\documentclass[12pt]{article}
\usepackage{graphicx}
\usepackage{epsfig}
\usepackage{epstopdf}
\usepackage{amsmath}
\usepackage{amsfonts}
\usepackage{amssymb}
\newcommand{\rom}[1]{\mathrm{#1}}

\newcommand{\beq}{\begin{equation}}
\newcommand{\eeq}{\end{equation}}
\newcommand{\be}{\begin{equation}}
\newcommand{\ee}{\end{equation}}
\newcommand{\beqa}{\begin{eqnarray}}
\newcommand{\eeqa}{\end{eqnarray}}
\newcommand{\beqar}{\begin{eqnarray*}}
\newcommand{\eeqar}{\end{eqnarray*}}
\newcommand{\bea}{\begin{eqnarray}}
\newcommand{\eea}{\end{eqnarray}}

\newcommand{\ka}{\kappa}
\newcommand{\eps}{\epsilon}
\newcommand{\bz}{\bar{z}}

\newcommand{\reef}[1]{(\ref{#1})}

\newcommand{\diag}{\textrm{diag}}

\numberwithin{equation}{section}

\begin{document}

\setlength{\unitlength}{1mm}

\begin{titlepage}

\begin{flushright}
MIT-CTP-3920
\end{flushright}
\vspace{1cm}

\begin{center}
{\bf \Large Bicycling Black Rings}
\end{center}

\vspace{7mm}

\begin{center}
Henriette Elvang$^{a}$ and  Maria J. Rodriguez$^{b,\; c}$

\vspace{.5cm}
{\small {\textit{$^{a}$Center for Theoretical Physics, }}\\
{\small \textit{Massachusetts
Institute of Technology, Cambridge, MA 02139, USA}}}\\
\vspace{2mm}
{\small \textit{$^{b}$Departament de F{\'\i}sica Fonamental}}\\
{\small \textit{Universitat de Barcelona, Diagonal 647, E-08028,
    Barcelona, Spain}} \\
\vspace{2mm}
 {\small \textit{$^{c}$Department of Physics}}\\
{\small \textit{University of California Santa Barbara, CA 93106, USA}}

\vspace*{0.5cm}
{\small {\tt elvang@lns.mit.edu, majo@ffn.ub.es}}
\end{center}

\vspace{1cm}

\begin{abstract}

We present detailed physics analyses of two different 4+1-dimensional asymptotically flat vacuum black hole solutions with spin in two independent planes: the doubly spinning black ring and the bicycling black ring system (``bi-rings''). The latter is a new solution describing two concentric orthogonal rotating black rings which we construct using the inverse scattering technique. We focus particularly on extremal zero-temperature limits of the solutions. We construct the phase diagram of currently known zero-temperature vacuum black hole solutions with a single event horizon, and discuss the non-uniqueness introduced by more exotic black hole configurations such as bi-rings and multi-ring saturns. 

\end{abstract}

\end{titlepage}

\tableofcontents

\newpage

\section{Introduction}

Black holes with zero temperature are of considerable interest since
one has a good chance of understanding the microscopic origin of their
physical properties, such as their entropy. Asymptotically flat supersymmetric black holes fall in this
class of solutions, and there has been considerable progress in
understanding their properties in string theory since the pioneering
work of Strominger and Vafa \cite{SV}. Non-supersymmetric extremal
black holes with zero temperature are likewise of interest. Recent
progress on understanding the microscopic nature of non-supersymmetric
zero temperature black holes includes \cite{EE,RE,EEF1,FL,EH,HR,Dias:2007nj}.

Non-supersymmetric black hole solutions are in general more difficult to work
with, not just in terms of understanding their microscopics, but also
because they are typically not easy to come by: exact solutions for
non-supersymmetric black holes tend to be harder to construct than
their supersymmetric cousins. This is in particular true for
multi-centered solutions. With their harmonic functions, supersymmetric black holes  can easily be superimposed, but non-supersymmetric systems can exhibit strong interactions
between the individual black hole components, and the solutions are consequently  
more involved.  

Application of integrability methods in higher-dimensional gravity has
recently allowed progress on construction of new exact black hole vacuum
solutions. For instance, the inverse scattering technique was used in
the construction of the first asymptotically flat multi-black hole
vacuum solution \cite{EF}. The solution --- named ``black saturn'' for
its characteristic appearance, a black ring balanced by rotation
around a spherical black hole --- exhibited clear signs of
interactions, including gravitational frame-dragging. Other novel
properties included a large degree of continuous non-uniqueness even
for zero total angular momentum  \cite{EF,EEF}. 
The black saturn system, as constructed in \cite{EF}, did not have a zero temperature limit. However, based on the results found in this paper, we propose the existence of an extremal zero-temperature black saturn solution, and we discuss the consequences of its expected  continuous non-uniqueness for the phase diagram of extremal zero-temperature vacuum black holes.

In this paper we study two different 4+1-dimensional black hole systems with spin in the two independent planes of rotation, and we pay attention particularly to
limits that give extremal zero temperature black hole configurations.

\begin{figure}[t!]
    \begin{center}
        \includegraphics[width=7cm]{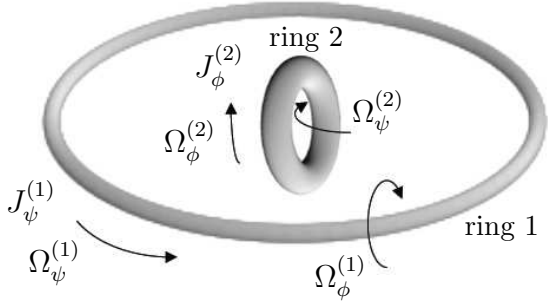}
    \end{center}
\begin{picture}(0,0)(0,0)
\put(50,10){$\Omega_\psi^{(1)}$}
\put(88,8){$\Omega_\phi^{(1)}$}
\put(93,30){$\Omega_\psi^{(2)}$}
\put(68,26){$\Omega_\phi^{(2)}$}
\put(47,18){$J_\psi^{(1)}$}
\put(72,36){$J_\phi^{(2)}$}
\put(108,15){\small ring 1}
\put(82,40){\small ring 2}
\end{picture}
\vspace{-5mm}
\caption{\small Bicycling black rings in orthogonal planes.}
    \label{fig:ringPic}
\end{figure}

The first of these systems is a new vacuum solution which we call ``bicycling
black rings'', or simply ``bi-rings'' for short. It is a balanced
configuration of two singly spinning concentric black rings placed in orthogonal planes. Singly spinning means that they each carry ``intrinsic'' angular momentum
only in the plane of the ring, i.e.~spin along the $S^1$ direction of the horizon. One
can think of the system as the superposition of two of the original
black rings of \cite{ER} in orthogonal planes. This is sketched in
figure \ref{fig:ringPic}. Note that the bi-ring solution is different from the so-called ``di-ring'' solutions \cite{IM,EK} where the two concentric rings lie in the same plane.

The bi-ring solution has three commuting Killing vectors (``$U(1)^3$
symmetry''): it is stationary, and the isometry of the $S^1$ of one
ring is the isometry of the azimuthal angle of the $S^2$ of the other
ring, and vice versa. We construct the solution using the inverse
scattering method.
The bi-ring solution exhibits 1-fold continuous
non-uniqueness after balance conditions have been imposed. This freedom
corresponds to distributing the total mass between the two black rings.

The two black rings in the bi-ring system interact with each
other. The rotation of the $S^1$ of one ring affects the $S^2$ of the
other by gravitational frame-dragging. This causes the $S^2$ to
rotate, so each ring has two non-vanishing angular velocities
corresponding to rotation in the two independent planes. The angular
momentum of the $S^2$ is effectively bounded by 
the 3+1-dimensional Kerr bound. The solution has a zero temperature
limit, and in order to understand what happens in this limit we study in detail another solution, namely the doubly spinning black ring. 

The doubly spinning black ring solution had long been anticipated. It
is a single black ring balanced by angular momentum in the plane of
the ring, but with angular momentum also in the orthogonal plane,
corresponding 
to rotation of the two-sphere.  
Kudoh \cite{K} found branches of this solution numerically, but the
true breakthrough\footnote{The unbalanced solution describing a black ring with
  rotation on the $S^2$ had been constructed by the inverse scattering
  method \cite{Mishima:2005id,Tomizawa:2005wv}, and also independently, 
  in a simpler form, in \cite{PF}.} 
 was Pomeransky and Sen'kov's construction of the exact doubly spinning
 black ring solution \cite{PS}. They used a clever implementation of
 the inverse scattering method, and moreover presented the solution in
 the more intuitive ring-type coordinates \cite{BRreview} rather than
 axisymmetric coordinates which are natural for the solution
 generating technique. A more general unbalanced doubly spinning black ring solution has also been found \cite{Morisawa:2007di}.

We provide in this paper a detailed analysis of the physical
properties of the Pomeransky-Sen'kov doubly spinning black
ring \cite{PS}. These results have not previously been presented in 
the literature. Our results include the structure of the phases of
doubly spinning black rings. We examine two different extremal
zero-temperature limits. One is shown to result in a regular
extremely spinning zero temperature black ring solution. The other limit
appears to be singular, but this is just a coordinate
singularity. A coordinate transformation shows that the resulting solution is nothing but the extremal doubly spinning Myers-Perry black hole, which is a regular black hole with zero temperature.

The zero-temperature limit of the bicycling black ring system is similar to
the second extremal limit of the doubly spinning black ring, and thus we find evidence that the rings in this limit collapse to the extremal Myers-Perry black hole.

Inspired by the analysis of the bi-ring system and the doubly spinning black ring, we speculate about the structure of the
phase diagram for zero-temperature black holes.
It should be emphasized that our bi-ring solution is not the most general one. The obvious generalization is constructed from two doubly spinning black rings in orthogonal planes.
In our discussion of the general phase diagram we consider the
possibilities of generalized bicycling black rings as well as their even more exotic
generalizations, which include multi-bicycles (tandems) and bi-ring saturns.
The richness of the phase structure of higher-dimensional black holes is striking,
even in a limit as specialized as that of zero-temperature.

Supersymmetric versions of saturns, di-rings, bi-rings and many-ring systems can be constructed via superposition of the harmonic functions that characterize the individual black hole components of the system. This was first done for rings in \cite{GG}. Supersymmetric saturns with the black hole moved off the plane of the ring were studied in \cite{Bena:2005zy}. A key point in our analysis of the non-supersymmetric multi-black hole solutions is to track the interaction between the black holes in the system; this is not possible in the supersymmetric configurations where the mutual BPS-ness cancels out interaction effects.

The organization of the paper is as follows. We construct and analyze
the bi-ring solution in section \ref{s:birings}. In
section \ref{s:sym} we study in detail a subfamily of the solutions
where the two orthogonal rings are identical. We compare the results with a
simple model obtained by superimposing two singly-spinning black rings
and neglecting interactions. The doubly spinning black ring of \cite{PS} is analyzed in
detail, and its physics discussed, in section \ref{s:doubly}. We
describe in section \ref{s:TzBike} the zero temperature scaling limit of the bi-ring
solution and argue that it corresponds to a collapse to a single extremal Myers-Perry black hole. We conclude with a discussion of zero temperature phases of
4+1-dimensional black holes in section \ref{s:disc}. Two appendices
are included: appendix \ref{app:hori} contains details of the horizon
metric for the bi-ring solution, and appendix \ref{app:MP} reviews
relevant properties of the Myers-Perry black hole \cite{MP}.

\vspace{3mm}
Note: while this work was in progress, the paper \cite{Izumi:2007qx} appeared, presenting also  an orthogonal two-ring solution.

\section{Bicycling black rings: Construction and analysis}
\label{s:birings}

\subsection{Construction}

The inverse scattering method was recently used to construct the black
saturn solution \cite{EF}. Our approach here is very similar, so we
keep the presentation brief. 
The construction takes as input a seed solution on which we perform a
series of ``soliton transformations'' and rescalings \cite{pom}. The
transformations introduce new parameters, called Belinsky-Zakharov
(BZ) parameters, which are organized in ``BZ vectors''. For
definitions and review of the method, see \cite{EF}. We also refer
  the reader to  \cite{BZ1,BZ2,BV} for literature on the inverse
  scattering technique and to \cite{ER2} and \cite{Harmark} for Weyl solution
  techniques and rod diagram representations.

\subsubsection{Seed solution}

\begin{figure}[t!]
\begin{picture}(0,0)(0,0)
\put(-4,22){$t$}
\put(-4,11){$\phi$}
\put(-4,0){$\psi$}
\put(7,-6){$a_1$}
\put(18,-6){$a_2$}
\put(30,-6){$a_3$}
\put(41,-6){$a_4$}
\put(52,-6){$a_5$}
\put(63.5,-6){$a_6$}
\put(75,-6){$a_7$}
\end{picture}
    \centering
        \includegraphics[width=8.5cm]{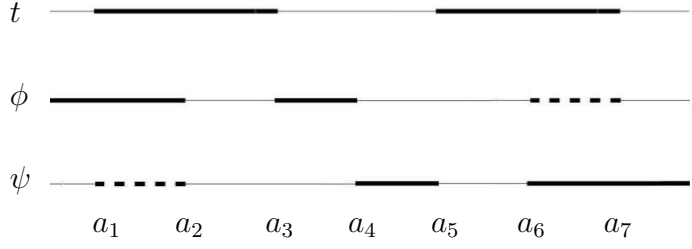}
        \bigskip
\caption{\small Rod configuration representing the sources for the seed
  metric $G_0$. Solid black lines in the figure correspond to rod
  sources of uniform  
  density $+1/2$ and the dashed rods to uniform densities  $-1/2$.}
    \label{fig:rods}
\end{figure}

The seed solution is represented by the rod diagram given in figure
\ref{fig:rods}. 
There are two negative density rods: one $[a_1,a_2]$ in
the $\psi$-direction and the other $[a_6,a_7]$ in the
$\phi$-direction. These are included to facilitate adding angular
momentum to each of the two black rings \cite{EF}.

The corresponding seed metric is given by
\bea
  ds^2 &=&  (G_0)_{ab} \, dx^a dx^b + e^{2\nu_0} \Big( d\rho^2 + dz^2\Big)
\eea
where $x^a = (t,\phi,\psi)$ and\footnote{An integration constant in $e^{2\nu_0}$ is set to one in order to make the final solution asymptotically flat.}
\bea
   \label{seed1}
   G_0 &=& \diag\left\{
   - \frac{\mu_1\, \mu_5}{\mu_3\,  \mu_7},\,
   \frac{\rho^2\,  \mu_3\,  \mu_7}{\mu_2 \, \mu_4 \, \mu_6},\,
   \frac{\mu_2\,  \mu_4\, \mu_6}{\mu_1\, \mu_5}
   \right\} \, , \\[2mm]
  \label{Hexp2nu}
  e^{2\nu_0}
   &=&  \frac{\mu_2\mu_4\mu_6}{\mu_1\mu_5}
         \frac{\left(\prod_{1 \le i < j \le 7}Z_{ij}\right)}
                { \Big[ Z_{15} Z_{26} Z_{37} Z_{24} Z_{46}\Big]^{3}
                  \left(\prod_{i=1}^7 Z_{ii}\right)} \, .
\eea
We use
\bea
  \label{themuis}
  \mu_i = \sqrt{\rho^2+(z-a_i)^2} - (z-a_i) \, ,
\eea
where the $a_i$ are the rod endpoints, and we have introduced
\bea
  \label{Zs}
  Z_{ij} = \rho^2 + \mu_i \mu_j \, .
\eea
Note that $\det{G_0}=-\rho^2$.  

We assume the ordering
\bea
  \label{order}
  a_1<a_2< a_3 < a_4 < a_5 <a_6<a_7 
\eea
of the rod endpoints.
When $a_6=a_7$ and $a_4=a_5$ the solution reduces to the
seed solution used to construct black saturn. With its naked singularities (due to the negative density rods) the seed solution is not by itself of physical interest. As shown in \cite{EF}, the soliton transformations which add angular momentum to the solution also make it possible to fully eliminate the naked singularities.

\subsubsection{The 2-soliton transformation}

We construct the bicycling black ring solution as follows (see \cite{EF} for details):

\begin{enumerate}
\item Perform the following two 1-soliton transformations on the
  seed solution \reef{seed1}:
\begin{itemize}
  \item Remove an anti-soliton at $z=a_1$ with trivial BZ vector
  (1,0,0); this is equivalent to dividing $(G_0)_{tt}$ by
  $-\rho^2/\bar{\mu}_1^2 =-\mu_1^2/\rho^2$.
  \item Remove a soliton at $z=a_7$ with trivial BZ vector (1,0,0);
    this is equivalent to dividing $(G_0)_{tt}$ by
    $\left(-\rho^2/\mu_7^2\right)$.
\end{itemize}
The result is the metric matrix
\bea
   G_0'=\diag\left\{
   -\frac{\mu_5\, \mu_7}{\mu_1\,  \mu_3},\,
   \frac{\rho^2\,  \mu_3\,  \mu_7}{\mu_2 \, \mu_4 \, \mu_6},\,
   \frac{\mu_2\,  \mu_4\, \mu_6}{\mu_1\, \mu_5}
   \right\} \, .
\eea
\item Rescale $G_0'$ by a factor of $-\frac{\mu_1}{\mu_7}$
  to find
\begin{equation}
  \tilde{G}_0=-\frac{\mu_1}{\mu_7} G_0'
  =\diag\left\{
   \frac{\mu_5}{\mu_3},\,
   \frac{\mu_1 \, \bar{\mu}_2\,  \mu_3}{\mu_4 \, \mu_6},\,
  - \frac{\mu_2\,\mu_4\, \mu_6}{\mu_5\, \mu_7}
   \right\} \, ,
\end{equation}
where $\bar\mu_2=-\rho^2/\mu_2$. This will be the seed for the next
soliton transformation.
\item The generating matrix can be found from  $\tilde{G}_0$. It is
\bea
  \tilde{\Psi}_0(\lambda,\rho,z)
  =\diag\left\{
   \frac{(\mu_5-\lambda)}{(\mu_3-\lambda)},\,
   \frac{(\mu_1-\lambda) \, (\bar{\mu}_2-\lambda)\,  (\mu_3-\lambda)}
   {(\mu_4-\lambda) \, (\mu_6-\lambda)},\,
  - \frac{(\mu_2-\lambda)\,(\mu_4-\lambda)\, (\mu_6-\lambda)}
  {(\mu_5-\lambda)\, (\mu_7-\lambda)}
   \right\} .~
\eea
Note $\tilde{\Psi}(0,\rho,z) = \tilde{G}_0$.
\item Perform now a 2-soliton transformation with $ \tilde{G}_0$ as
  seed:
\begin{itemize}
\item Add an anti-soliton at $z=a_1$ (pole at $\lambda=\bar\mu_1$)
 with BZ vector
 $m_0^{(1)} =(1,0,c_1)$, and
\item Add a soliton at  $z=a_7$ (pole at $\lambda=\mu_7$) with BZ vector
 $m_0^{(2)} =(1,b_2,0)$.
 \end{itemize}
Denote the resulting metric $\tilde{G}$. The constants $c_1$ and $b_2$
are the BZ parameters of the transformation.
\item Rescale $\tilde{G}$ to find
\bea
  G = -\frac{\mu_7}{\mu_1}  \tilde{G} \, .
\eea
This is needed to undo the rescaling of step 2, so that $\det G =
-\rho^2$.
\item The metric factor $e^{2\nu}$ is constructed using eq.~(2.14) of \cite{EF}. 
\end{enumerate}

\noindent The result $(G, e^{2\nu})$ is the final solution.


\subsubsection{Solution}

The bicycling black ring solution can be written as
\bea
  \nonumber
  ds^2 &=& - \frac{H_y}{H_x} \Big[ dt - \frac{\omega_\phi}{H_y} \, d\phi
      - \frac{\omega_\psi}{H_y} \, d\psi  \Big]^2
      + H_y^{-1} \Big[
         G_x \, d\phi^2
        +G_y \, d\psi^2
        -2 J_{xy}\, d\phi \, d\psi \Big] + P \,  H_x  \Big[ d\rho^2 + dz^2 \Big] \, .
     \label{sol}
\eea
We have written $e^{2\nu} = P \,  H_x$. 
The metric is given in terms of the functions:\vspace{1mm}
\bea
  P = \frac{\mu_2 \, Z_{23} \, Z_{25} \, Z_{34} \,  Z_{35} \, Z_{36}
    \, Z_{45} \, Z_{47} \, Z_{56} \, Z_{57} \, Z_{67}}
   {\mu_1\,  \mu_5^4\,  \mu_7 \, (\mu_3 - \mu_7)^4
    \, Z_{12} \, Z_{13} \, Z_{14} \, Z_{15}^2
    \, Z_{16} \, Z_{17} \, Z_{24}^2 \, Z_{26}^2 \, Z_{27}
    \, Z_{37}^2 \, Z_{46}^2 \,
    \Big[\prod_{i=1}^7 Z_{ii} \Big] \, , } \, ,
\eea
(the $Z_{ij}$ are defined in \reef{Zs}),
\bea
    H_x &=&
    \Big( M_0 + c_1^2  \, M_1 + b_2^2
      \, M_2 - c_1^2  \, b_2^2 \,  M_3 \Big) \, , \\[3mm]
    H_y &=& \frac{\mu_5}{\mu_3}
    \Big( \frac{\mu_1}{\mu_7} M_0
    - c_1^2  \, \frac{\rho^2}{\mu_1\,\mu_7} M_1
    - b_2^2  \, \frac{\mu_1\,\mu_7}{\rho^2} M_2
    - c_1^2  \, b_2^2 \,   \frac{\mu_7}{\mu_1} M_3 \Big) \, ,\\[3mm]
    G_x &=&
    \frac{\mu_1\, \mu_5 \, \rho^2}{\mu_2\,\mu_4\,\mu_6}
    \Big( M_0
    - c_1^2  \,\, \frac{\rho^2}{\mu_1^2}\, M_1
    + b_2^2  \, M_2
    + c_1^2  \, b_2^2 \,\,  \frac{\rho^2}{\mu_1^2} \, M_3 \Big) \, ,
    \\[3mm]
    G_y &=&
    \frac{\mu_2\,\mu_4\,\mu_6}{\mu_3\, \mu_7}
    \Big( M_0
    + c_1^2  \, M_1
    - b_2^2  \,\, \frac{\mu_7^2}{\rho^2}\, M_2
    + c_1^2  \, b_2^2 \,\, \frac{\mu_7^2}{\rho^2}\, M_3 \Big) \, ,
\eea
and
\bea
  \nonumber
  J_{xy} &=& c_1 \, b_2 \;
     \rho^2 \, \mu_1 \mu_2 \, \mu_3 \, \mu_4 \, \mu_5^2 \, \mu_6 \,
       (\mu_3 - \mu_7)^2 (\mu_4 - \mu_7)
       (\mu_5 - \mu_7) (\mu_6 - \mu_7) \\[2mm]
    && \hspace{5mm} \times
       \, Z_{11} \, Z_{77} \, Z_{12} \, Z_{13} \, Z_{14}
         \, Z_{15}^2 \, Z_{16} \, Z_{17} \, Z_{27} \, ,
\eea
with
\bea
  M_0 &=&  \mu_4 \, \mu_5^3 \, \mu_6 \, \mu_7 \,  (\mu_3 - \mu_7)^4
   \, Z_{12}^2 \, Z_{13}^2 \, Z_{14}^2 \, Z_{16}^2
    \, Z_{17}^2 \, Z_{27}^2  \, , \\[2.5mm]
  M_1 &=& \rho^2\, \mu_1^2 \, \mu_2 \, \mu_3\, \mu_4^2 \,
     \mu_5 \, \mu_6^2 \, (\mu_1 - \mu_7)^2 \,  (\mu_3 - \mu_7)^4
  \, Z_{15}^4 \, Z_{17}^2 \, Z_{27}^2 \, , \\[2.5mm]
  M_2 &=& \rho^4\, \mu_1 \, \mu_2 \, \mu_3^2 \, \mu_5^2 \, \mu_7  \,
     (\mu_4 - \mu_7)^2  \, (\mu_5 - \mu_7)^2 \,  (\mu_6 - \mu_7)^2
  \, Z_{12}^2 \, Z_{13}^2 \, Z_{14}^2 \, Z_{16}^2\, , \\[2.5mm]
  M_3 &=& \rho^4\, \mu_1^3 \, \mu_2^2 \, \mu_3^3 \, \mu_4 \, \mu_6  \,
     (\mu_4 - \mu_7)^2  \, (\mu_5 - \mu_7)^2 \,  (\mu_6 - \mu_7)^2
  \, Z_{15}^4 \, Z_{17}^2 \, .
\eea
Finally we also have
\bea
  \omega_\psi &=&
  c_1  \, \frac{Z_{11}}{\mu_1}
  \sqrt{\frac{\mu_2 \, \mu_4 \, \mu_6}{\mu_3 \, \mu_7 \, \rho^2}}
  \Big( \sqrt{M_0\, M_1}
  + b_2^2 \, \frac{\mu_7}{\rho} \sqrt{M_2 \, M_3}  \Big)
  \, , \\[2mm]
  \omega_\phi &=&
  b_2  \, \frac{Z_{77}}{\mu_7}
  \sqrt{\frac{\mu_1 \, \mu_5}{\mu_2 \, \mu_4 \, \mu_6}}
  \Big( \sqrt{M_0\, M_2}
  + c_1^2 \, \frac{\rho}{\mu_1} \sqrt{M_1 \, M_3}  \Big) \, .
\eea

Note that changing the sign of the BZ parameter $c_1$, $c_1 \to - c_1$
is equivalent to reversing the direction of rotation by taking $\psi
\to - \psi$. Likewise, $b_2 \to -b_2$ simply corresponds to taking $\phi
\to - \phi$. Thus we choose $c_1$ and $b_2$ to be negative
without loss of generality. This choice gives positive angular momenta and angular velocities.


\subsection{Analysis}

\subsubsection{Parameterization}
The solution is parameterized by the rod endpoints $a_i$ and the two BZ parameters $c_1$ and $b_2$. The parameters are all dimensionful. It is convenient to introduce an overall scale $L$, 
\bea
  L^2 = a_7-a_1 \; ,
\eea
and then express the solution in terms of dimensionless parameters $\ka_i$ defined as
 \bea
  \kappa_{i} = \frac{a_{i+1} - a_1}{L^2}
 \eea
for $i=1,2,...,5$. The ordering \reef{order} implies that
\bea
  \label{kaorder}
  0 \le \ka_1 \le \ka_2\le \ka_3 \le \ka_4 \le \ka_ 5 \le  1\;.
\eea
It is also useful to shift the $z$ coordinate as
\bea
 z = L^2 \bar{z} +a_1 \;,
\eea
so that $\bar{z}$ is dimensionless.

An analysis of the metric components $G_{ab}$, $a,b=t,\psi,\phi$, near the rod endpoints shows that there are divergences near $\bar{z}=0$  and $\bar{z}=1$, i.e.~near $z=a_1$ and $z=a_7$. Just as for the black saturn solution, these singularities can be eliminated by an appropriate choice of the BZ parameters. This fixes $c_1$ and $b_2$ to be\footnote{Changing the sign of the BZ parameters corresponds to reversing the sense of rotating of the rings, so without loss of generality we have assumed  $c_1$ and $b_2$ to be negative.}
\bea
  \label{thec1b2}
  c_1 \; =\; 
  -\frac{L\sqrt{\;2 \; \kappa_1 \; \kappa_2 \; \kappa_3 \; \kappa_5}}
  {\kappa_4} \, ,
  \hspace{8mm}
  b_2\;  =\;  -L\;(1-\kappa_2)\sqrt{  \frac{2\;(1-\kappa_1)}
     {(1-\kappa_3) \;(1- \kappa_4) \; (1-\kappa_5)}} \;.
\eea

In what follows we will always impose the smoothness conditions \reef{thec1b2}. The $\rho=0$ metric is then smooth across  $\bar{z}=0$  and $\bar{z}=1$, and there are no signs of the negative density rods of the seed solution.
We describe the rod configuration of the bicycling black ring solution in the next subsection. 

With $c_1$ and $b_2$ fixed by \reef{thec1b2}, the full solution is parametrized by the five  $\ka_i$-parameters, subject to the ordering \reef{kaorder}, and the scale $L$. 


\subsubsection{Rod structure}

\begin{figure}[t!]
\begin{center}
        \includegraphics[width=8.5cm]{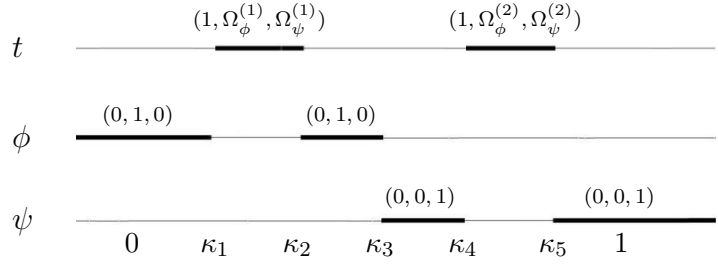}
\end{center}
\begin{picture}(0,0)(0,0)%
\put(34,32){$t$}%
\put(34,20){$\phi$}%
\put(34,9){$\psi$}%
\put(49,6){\small $0$}%
\put(59,6){\small $\ka_1$}%
\put(70,6){\small $\ka_2$}%
\put(81,6){\small $\ka_3$}%
\put(92,6){\small $\ka_4$}%
\put(104,6){\small $\ka_5$}%
\put(114,6){\small $1$}%
\put(58,36){\scriptsize$(1,\Omega_\phi^{(1)},\Omega_\psi^{(1)})$}%
\put(92,36){\scriptsize$(1,\Omega_\phi^{(2)},\Omega_\psi^{(2)})$}%
\put(46,23.5){\scriptsize$(0,1,0)$}%
\put(73,23.5){\scriptsize$(0,1,0)$}%
\put(83.5,12.5){\scriptsize$(0,0,1)$}%
\put(110,12.5){\scriptsize$(0,0,1)$}%
\end{picture}
    \caption{\small
Rod structure of the regular bicycling black ring solution. 
Rod directions are shown over each rod.}
    \label{fig:rodsdirection}
\end{figure}

Imposing  \reef{thec1b2} the
rod structure can be summarized as (see figure \ref{fig:rodsdirection}):
\begin{itemize}
\item The semi-infinite rod $]-\infty,\ka_1]$ and the finite rod
$[\ka_2,\ka_3]$ have direction $(0,1,0)$.

\item The semi-infinite rod $[\ka_5,\infty[$ and the finite rod
    $[\ka_3, \ka_4]$ have direction $(0,0,1)$.

\item The finite rod $[\ka_1, \ka_2]$ corresponds to a black ring whose
  horizon has the $S^1$ parameterized by $\psi$ and the $S^2$ by
  $(z,\phi)$. This rod has direction
  $(1,\Omega_\phi^{(1)},\Omega_\psi^{(1)})$.

  Likewise, the finite rod $[\ka_4, \ka_5]$ corresponds to a black ring whose
  horizon has the $S^1$ parameterized by $\phi$ and the $S^2$ by
  $(z,\psi)$. This rod has direction
  $(1,\Omega_\phi^{(2)},\Omega_\psi^{(2)})$.

\end{itemize}

The angular velocities are
\bea
  &&\Omega^{(1)}_\phi \; =\; \frac{1}{L\, (1-\ka_2)}
  \sqrt{\frac{(1-\ka_3)(1-\ka_4)(1-\ka_5)}{2\, (1-\ka_1)}} \, ,
  \hspace{8mm}
  \Omega^{(1)}_\psi \; =\;  \frac{\ka_4}{L}
  \sqrt{\frac{\ka_1}{2\, \ka_2\,\ka_3\,\ka_5}} \, , \\[3mm]
  &&\Omega^{(2)}_\phi \; =\;  \frac{1-\ka_2}{L}
  \sqrt{\frac{(1-\ka_5)}{2\, (1-\ka_1)(1-\ka_3)(1-\ka_4)}} \, ,
  \hspace{1.2cm}
  \Omega^{(2)}_\psi \; =\;  \frac{1}{L\, \ka_4}
  \sqrt{\frac{\ka_1 \,\ka_2 \, \ka_3}{2\,  \ka_5}} \, .~~~~~~~
\eea
We use superscripts $(i)$, $i=1,2$, for the quantities associated with ring 1 (horizon located at $\rho = 0$ and $\ka_1 \le z \le \ka_2$)
and ring 2 (horizon at $\rho = 0$ and $\ka_4 \le z \le \ka_5$).

\subsubsection*{Exchange symmetry}

The rod picture figure \ref{fig:rodsdirection} suggests that the solution has a symmetry corresponding to interchanging the two black rings. Indeed we have confirmed that the transformation
\bea
  \nonumber
  &&
  \ka_1 \to 1-\ka_5\, ,~~~
  \ka_2 \to 1-\ka_4\, ,~~~
  \ka_3 \to 1-\ka_3\, ,~~~
  \ka_4 \to 1-\ka_2\, ,~~~
  \ka_5 \to 1-\ka_1\, ,~~~ \\[2mm]
  \nonumber
  && \bz \to 1-\bz \, ,~~~~~~
     \psi \leftrightarrow \phi \, ,
\eea
exchanges the physical parameters of the two black rings, as well as
their balance conditions.\footnote{This is only true when the conditions \reef{thec1b2} are imposed; if not imposed, one must in addition transform $c_1$ and $b_2$.}


\subsubsection{Asymptotics and balance}
The solution \reef{sol} is asymptotically flat. 
To see this, introduce asymptotic coordinates $(r,\theta)$ as
$\rho=\frac{1}{2} \,r^2 \sin {2\theta}$ and
$z=\frac{1}{2}\, r^2 \cos {2\theta}$.
In the asymptotic limit $r \to \infty$, the metric then approaches
\be\label{flatmetric}
  ds^2=-dt^2+dr^2
  +r^2(d\theta^2+\sin^2 \theta\, d\psi^2+\cos^2 \theta\, d\phi^2) \, .
\ee
Regularity conditions\footnote{The periodicities \reef{period} follow from regularity conditions on the rods $]-\infty,\kappa_1]$  and $[\kappa_5,\infty[$. The general regularity condition is that when $G_{\phi\phi} \to 0$ as $\rho \to 0$ then $\Delta\phi=2\pi\,\lim_{\rho\rightarrow0}
  \sqrt{\frac{\rho^2\, G_{\rho\rho}}{G_{\phi\phi}}}\label{phicond}$.
  Likewise for $\psi$.}
require the angular coordinates $\phi$ and $\psi$ have periods $2\pi$,
\be\label{period}
  \Delta\phi=2\pi\, ,~~~~~~
  \Delta\psi=2\pi \, ,
\ee
so indeed the metric is asymptotically flat.

Regularity on the rods  $[\kappa_2,\kappa_3]$ and $[\kappa_3,\kappa_4]$ requires non-trivial conditions, 
\beqa
  \label{bal1}
  1 ~=~ \frac{\Delta\phi}{2\pi} &=& \frac{\sqrt{\kappa_3\kappa_5(1-\kappa_1)
    (\kappa_3-\kappa_2)(\kappa_4-\kappa_1)(\kappa_4-\kappa_2)
    (\kappa_5-\kappa_2)}}{\kappa_4(1-\kappa_2)(\kappa_3-\kappa_1)
    (\kappa_5-\kappa_1)}\, ,  \\[2mm]
   \label{bal2}
  1~=~\frac{\Delta\psi}{2\pi} &=&\frac{\sqrt{\kappa_5(1-\kappa_1)(1-\kappa_3)
    (\kappa_4-\kappa_1)(\kappa_4-\kappa_3)(\kappa_4-\kappa_2)
    (\kappa_5-\kappa_2)}}{\kappa_4(1-\kappa_2)(\kappa_5-\kappa_1)
    (\kappa_5-\kappa_3)}\, .
\eeqa
These are the conditions for balancing each of the two black rings. If not imposed, then there are disks of conical singularities inside the rings. When studying the physical properties of the bicycling ring system, we always impose both balance conditions.

\subsubsection*{Solving the balance conditions}
We briefly outline the strategy for solving the balance conditions in practical applications. This is used in section \ref{s:TzBike}.
The squared ratio of \reef{bal1} and \reef{bal2} is
\bea
  1 = \frac{\ka_3\, (\ka_3-\ka_2)(\ka_5-\ka_3)^2}
      {(1-\ka_3)(\ka_4-\ka_3)(\ka_3-\ka_1)^2} \, .
\eea
This condition is linear in $\ka_2$ and $\ka_4$;
solving for $\ka_4$ gives
\bea
  \label{ka4}
  \ka_4=\ka_4^* \equiv \ka_3 + \frac{\ka_3 (\ka_3-\ka_2)(\ka_5-\ka_3)^2}
   {(1-\ka_3)(\ka_3-\ka_1)^2} \, .
\eea
Clearly  $\ka_3 \le \ka_4^*$, but we must also require that $\ka_4^* \le \ka_5$, and
that restricts the parameters on the RHS of \reef{ka4}. 
The simplest way to express this is as an upper bound on $\ka_5$,
\bea
  \ka_5 \le \ka_{5\rom{MAX}}(\ka_1, \ka_2, \ka_3)
  \equiv
  \frac{\ka_1^2\, (1-\ka_3) - 2\, \ka_1 \, \ka_3 \, (1-\ka_3)+\ka_3^2\, (1-\ka_2)}
  {\ka_3\, (\ka_3-\ka_2)} \, .
\eea
This condition ensures $\ka_3 \le \ka_4^* \le \ka_5$.

Plugging \reef{ka4} into either of the balance conditions \reef{bal1}
or \reef{bal2} we obtain a polynomial equation which is 8th order in
$\ka_1$, 4th order in $\ka_2$, 7th order in $\ka_3$, and 6th order in
$\ka_5$. Solving this condition for $\ka_2$ seems to be the simplest.
As a fourth order polynomial, one can obtain the roots
analytically, but in applications we will simply solve numerically for $\ka_2$ in order to impose balance.
For given $\ka_1$, $\ka_3$, and $\ka_5$, we select the real root(s)
$\ka_2^*$ which satisfy (a) $\ka_1 < \ka_2^* < \ka_3$, and (b) $\ka_5 \le \ka_{5\rom{MAX}}(\ka_1, \ka_2^*, \ka_3)$.

Imposing both balance conditions leaves the bi-ring solution with three dimensionless parameters $0<\ka_1 < \ka_3 < \ka_5<1$ and the scale $L$. Fixing the total ADM mass fixes the scale. Fixing further the only other conserved quantities, namely the two angular momenta, leaves a single free parameter. This continuous non-uniqueness parameter corresponds to the freedom of distributing the total mass between the two black rings.


\subsubsection{Physical parameters}
The ADM mass and angular momenta are
\beqa
  \label{mass}
  M&=&\frac{3\pi\, L^2}{4 G_5}(1+\kappa_2 -\kappa_4)\, , \\
  J_{\psi} &=&  \frac{\pi\,L^3}{\sqrt{2}\, G_5}
    \frac{\sqrt{\ka_1\, \ka_2 \, \ka_3 \, \ka_5}}{\ka_4}\, , \\
 J_{\phi} &=&   \frac{\pi\,L^3}{\sqrt{2}\, G_5}
    \frac{\sqrt{(1-\kappa_1)(1-\kappa_3)(1-\kappa_4 )(1-\kappa_5 )}}
    {(1-\kappa_2)}\, .
\eeqa
The mass is always positive.

We analyze the black ring horizons in appendix \ref{app:hori}. The horizon area and temperature of each black ring are found to be
\bea
  A_\rom{H}^{(1)} &=& 
   (2\pi)^2\,  L^3 \, (\ka_2-\ka_1)\, \frac{\sqrt{2\, \ka_2\, \ka_3\, \ka_5\,
    (1-\ka_1)(\ka_2-\ka_1)(\ka_4-\ka_1)}}
       {\ka_4 \, (\ka_3-\ka_1)(\ka_5-\ka_1)} \, , \\[2mm]
     A_\rom{H}^{(2)} &=&
      (2\pi)^2\,  L^3 \, (\ka_5-\ka_4)\, 
      \frac{\sqrt{2\, \ka_5\, (1-\ka_1)(1-\ka_3)(1-\ka_4)
       (\ka_5-\ka_4)(\ka_5-\ka_2)}}
       {(1-\ka_2)(\ka_5-\ka_1)(\ka_5-\ka_3)} \, ,
\eea
and
\bea
  T_\rom{H}^{(1)} &=& 
  \frac{\ka_4 \, (\ka_3-\ka_1)(\ka_5-\ka_1)} {2\pi L\sqrt{2\, \ka_2\, \ka_3\, \ka_5\,
    (1-\ka_1)(\ka_2-\ka_1)(\ka_4-\ka_1)}}\, ,\\[2mm]
   T_\rom{H}^{(2)} &=& 
   \frac{(1-\ka_2)(\ka_5-\ka_1)(\ka_5-\ka_3)}
  {2\pi L \sqrt{2 \, \ka_5\, (1-\ka_1)(1-\ka_3)(1-\ka_4)
   (\ka_5-\ka_2)(\ka_5-\ka_4)}} \, .
   \eea


\subsubsection*{Smarr relations}
The bicycling rings satisfy the Smarr relation
\bea
  \frac{2}{3} M = T_1 S_1 + T_2 S_2 + \Omega_\psi^{(1)} \, J_\psi
      + \Omega_\phi^{(2)} \, J_\phi \, ,
\eea
where $S^{(i)} = A_\rom{H}^{(i)}/(4G)$. This holds independently of the balance
conditions.

Komar integrals for the masses and angular momenta give
\bea
 M^{(1)} = \frac{3 \pi L^2}{4 G} \, \ka_2 \, , ~~~~
 M^{(2)} = \frac{3 \pi L^2}{4 G} \, (1- \ka_4) \, .
\eea
and 
\bea
  &&J_\psi^{(1)} = J_\psi \, ,~~~~~
  J_\phi^{(1)} = 0 \, ,~~~~~
  J_\phi^{(2)} = J_\phi \, ,~~~~~
  J_\psi^{(2)} = 0 \, .
\eea
The Komar masses and angular momenta add up to the total ADM mass and ADM angular momenta, and they satisfy
$M^{(i)} = \frac{3}{2}
  \big( T_\rom{H}^{(i)} S^{(i)} + \Omega_\psi^{(i)} \, J^{(i)}_\psi \big)$. This holds independently of the balance conditions, and should merely be viewed as a calculational check (see \cite{EEF}) of our results for the physical parameters.


\subsubsection{CTCs}

It can analytically be shown that there are no closed timelike curves
(CTCs) in the plane inside each rings. Outside the ring, in the plane
of the ring, we have checked numerically near the ring and found no
CTCs. We have also performed numerical checks in the bulk (off the
planes of the rings), and no CTCs where found. Our checks are
not exhaustive, and further progress may require a better coordinate
system. We do not expect the solution to have naked CTCs, in
particular we note that typical signs of CTCs are absent here, for
instance the horizon area remains positive and well-defined for all
admissible parameters.


\section{Symmetric bicycles}
\label{s:sym}

Our bi-ring solution is described by a single scale and five
dimensionless parameters, subject to two balance conditions. Thus for
fixed ADM mass, there are three independent parameters for the
balanced system. Asymptotically, the total angular momenta in
the two planes fix two of these parameters. Thus the balanced bi-rings
have 1-fold continuous non-uniqueness. This freedom corresponds to
continuously distributing the total mass between the two black rings. 

A particularly simple subclass of bi-ring solutions is obtained by
requiring that the two black rings are identical; i.e. that they have
the same temperatures, the same areas, and the same $S^1$ and $S^2$ angular velocities. We study this class of solutions in this section. 

Please note that in this paper we do not review properties of singly spinning black rings and their thin and fat ring branches in the phase diagram. This has by now been explained in much detail in several papers. We refer in particular to \cite{EEV} and the review \cite{BRreview}.

\subsection{A model}
Before examining in detail the symmetric bi-ring configuration, it is
useful to first model the system by superimposing
two singly spinning rings in orthogonal planes. This of course ignores
the interactions between the two rings, the significance of which we
soon discover. A similar model for black saturn was studied in \cite{EEF}.

\begin{figure}
\begin{center}
  \includegraphics{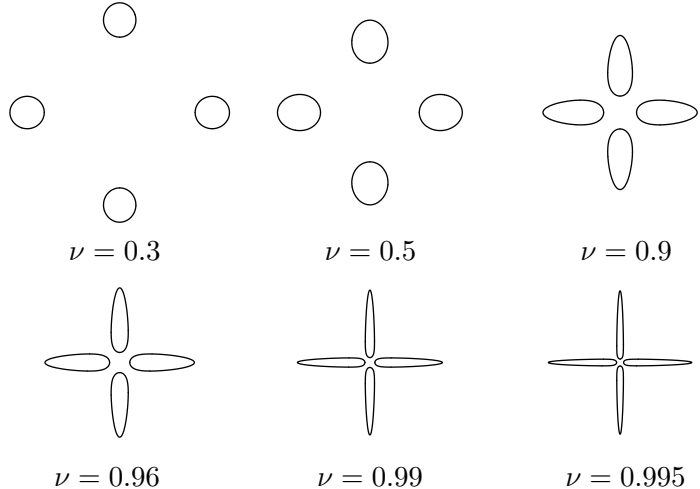}
\end{center}
\begin{picture}(0,0)(0,0)%
\put(45,40){\small $\nu=0.3$}%
\put(79,40){\small $\nu=0.5$}%
\put(113,40){\small $\nu=0.9$}%
\put(43,10){\small $\nu=0.96$}%
\put(78,10){\small $\nu=0.99$}%
\put(111,10){\small $\nu=0.995$}%
\end{picture}
\vspace{-8mm}
\caption{\small Visualization of the symmetric bicycling black ring
  system. Shown is the superposition of two identical 
  singly spinning rings in orthogonal planes for six different values of
  the angular momentum $j_\psi=j_\phi=\sqrt{(1+\nu)^3/(8\nu)}$ (recall \cite{EEF} that $0 <\nu < 1/2$ for the thin ring branch and $1/2 <\nu < 1$ on the fat black
  ring branch). The total mass is fixed to be the 
  same for each plot. The embeddings are plotted on the same scale. 
  Interactions between the two
  rings are ignored in this model, but in the real bi-ring solution
  interactions play an important role.}
     \label{fig:toyplot}
\end{figure}

The first purpose of the model is to illustrate the geometry of the 
2-ring configuration.   
This is done for six different values of the angular momentum in
figure \ref{fig:toyplot}. In each case we have plotted, for two
identical copies of singly spinning black ring, the isometric
embeddings of the diametrical cross section of the ring $S^2$'s, separated by
the inner horizon radius. These quantities, as well as further details
on black ring geometry and isometric embeddings, can be found in
\cite{EEV}. 

{}Black rings on the thin ring branch have two-spheres which are nearly
spherical and largely separated. As the angular momentum decreases so
does the inner horizon radius, and the $S^2$ of the horizon deforms as
shown in figure \ref{fig:toyplot}. For the singly spinning black ring,
the inner radius can become arbitrarily small, and as it goes to zero
the solution becomes nakedly singular. In the model the inner horizon
radius can also take arbitrarily small values, but this of course ignores
interactions between the two rings.  

The second purpose of the 2-ring model is to examine what we may encounter in
the real bi-ring system.
We introduce the usual fixed mass dimensionless variables 
\be 
  \label{jaH}
  j^2=\frac{27 \, \pi}{32\, G} \frac{J^2}{M^3} \, , ~~~~~~~
 a_\rom{H}=\frac{3}{16}\sqrt{\frac{3}{\pi}} \frac{A_\rom{H}}{(G \,M)^{3/2}} \,
\ee
for angular momentum and horizon area. 
In order to balance itself
against collapse, a singly spinning black ring must carry a certain
amount of angular momentum for the given mass. This gives a lower
bound on $j$, namely $27/32 \le j^2$. At $j^2 = 27/32$, the ring has maximum
horizon area, $a_\rom{H}=1$. The singly spinning ring can rotate with
arbitrarily high angular momentum, and as it does so the area goes to zero as $a_\rom{H} \sim (\sqrt{2}\, j)^{-1}$ for $j \to \infty$. 

We normalize the physical parameters for the bi-rings in terms of
the total mass. For the symmetric configuration, this is $2M$, with
$M$ the mass of each ring. The total area is $2 A_\rom{H}$. 
In the limit where the rings are large and thin, they have little interaction
and we expect the model to describe the system well. Using \reef{jaH} with $2M$ as the total mass we find that for the symmetric bi-ring model, the total horizon area goes as
$a_\rom{H}^\rom{total} \sim (4\sqrt{2}\, j)^{-1}$ for large $j$. This behavior is verified for the bi-ring solution in the next subsection. 

Ignoring interactions we likewise estimate the bi-ring to have
maximal area $a_\rom{H}^\rom{max}=2/2^{3/2} = 1/\sqrt{2} \approx 0.7$
and minimum angular momentum $j_\rom{min}^2=1/2^3(27/32) \approx
0.11$. We show below that the actual values
are $a_\rom{H}^\rom{max} \approx 0.485$ and  $j_\rom{min}^2 \approx
0.246$ for the bi-ring solution. This indicates that interaction effects are important for black rings near the cusp where the rings in the bi-ring system are closer to each other.

A clear sign of interactions in the true bi-ring configuration is that
the rotation of the $S^1$ of one ring drags the $S^2$ of the other
ring into rotation. We study this in the next subsection.

\subsection{The symmetric bi-ring phase}
In the symmetric bi-ring configuration the two rings are identical:
they have the same area and temperature, and the magnitudes 
of the angular momenta in the two planes of rotation are the same. 
It is easily verified that this is
obtained from a symmetric rod configuration with 
\bea
  \ka_5 = 1-\ka_1\,,~~~~~
  \ka_4 = 1-\ka_2\,,~~~~~
  \ka_3 = \frac{1}{2}\, .
\eea

The two balance conditions \reef{bal1}-\reef{bal2} are also identical, and an equilibrium
configuration is is therefore obtained by imposing a single condition,  
\bea
  \label{SYMbalance}
  1 = \frac{(1-\ka_1)(1-\ka_1-\ka_2)(1-2\ka_2)}{(1-\ka_2)^2(1-2\ka_1)^2} \, .
\eea
The parameters must satisfy
\bea
   0 < \ka_1 < \ka_2 < \frac{1}{2} \, ,
\eea
and this selects one solution $\ka_2^*$ to the balance condition \reef{SYMbalance}:
\bea
  \ka_2^* = \frac{1+3\ka_1-6\ka_1^2 - (1-2\ka_1)\sqrt{1+2\ka_1 - 3\ka_1^2}}
  {2(1+2\ka_1-4\ka_1^2)} \, .
\eea
Thus the balanced symmetric bi-ring system is parameterized by a
single parameter,  $\ka_1$. This is equivalent to a single balanced
black ring which is parameterized by a single parameter. 

\begin{figure} 
\begin{picture}(0,0)(0,0)
\put(-3,48){\small $a_\rom{H}^\rom{total}$}%
\put(-7,41){\small $2\sqrt{2}$}%
\put(-4,29.5){\small $2$}%
\put(-4,14){\small $1$}%
\put(55,-1){\small $j^2$}%
\put(36,-5){\small $\frac{1}{4}$}%
\put(-1,-4){\small $0$}%
\put(25,38){\small{MP BH}}
\put(15,4){\small{sym bi-ring}}
\put(68,47){\small $\omega$}
\put(147,-1){\small $r_\rom{inner}$}
\put(65.5,38){\small $1$}
\put(62,18){\small $0.5$}
\put(88,-5){\small $r_\rom{m}$}
\put(93,-5){\small $1$}
\put(116.5,-5){\small $1.5$}
\put(104,35){\small{$\omega_{S^1}$}}
\put(104,12){\small{$\omega_{S^2}$}}
\put(18,-10){\footnotesize Figure (a)}
\put(97,-10){\footnotesize Figure (b)}
\end{picture}
    \centering
          \includegraphics[height=4.5cm]{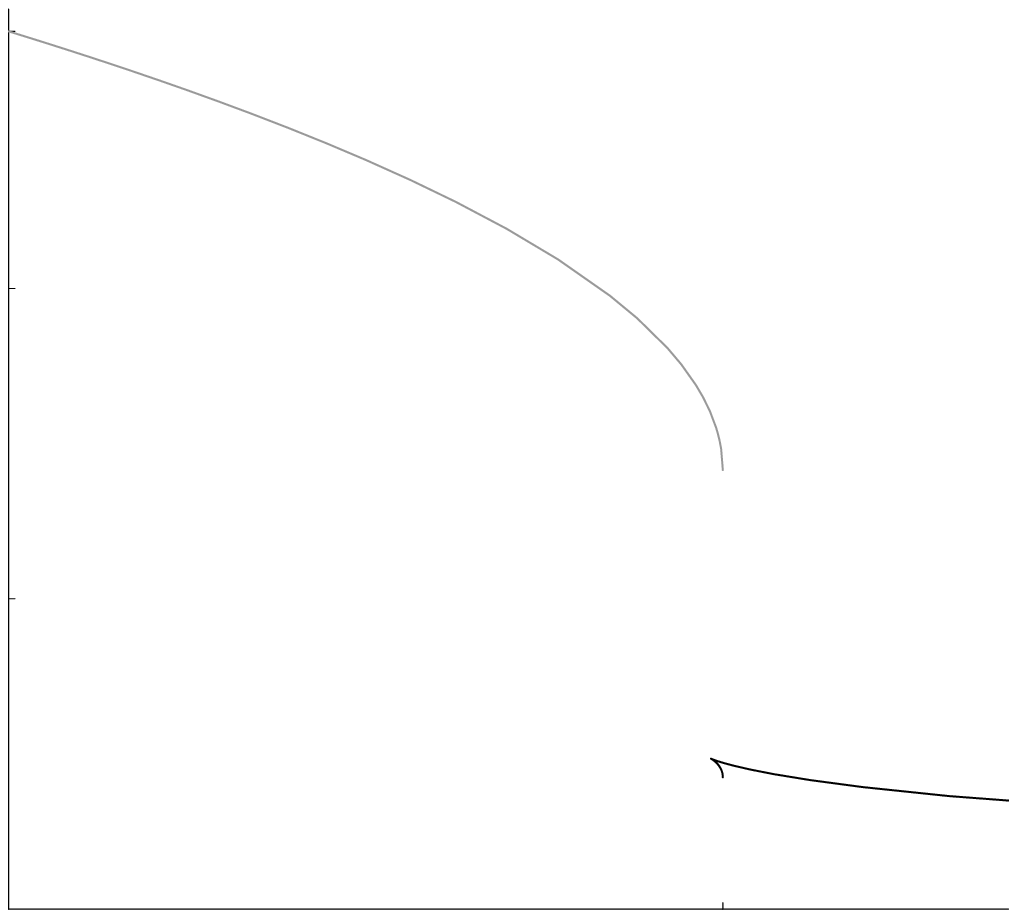}
          \hspace{1.4cm}
          \raisebox{-0.38cm}
          {\includegraphics[height=4.8cm]{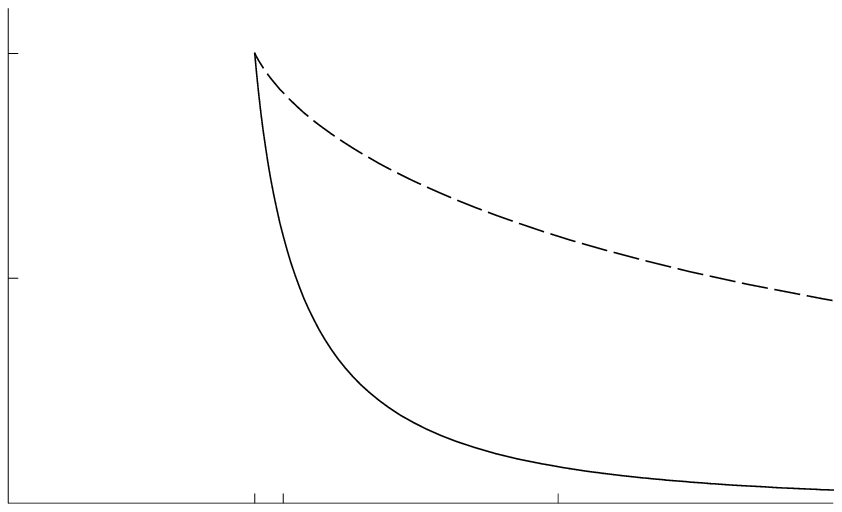}}
        \vspace{8mm}
    \caption{\small (a) Area $a_\rom{H}$ vs. $j^2$ for solutions with equal
    angular momenta in the two planes of rotation, $j \equiv j_\psi =
    j_\phi$. The \emph{gray} curve shows the Myers-Perry black hole
    phase and the \emph{black} curve is the symmetric bi-ring
    configuration. Both curves have endpoints at $j^2 = 1/4$. The
    solutions at the endpoints have finite area and zero temperature.
    (b) Angular velocities vs.~inner horizon radius $r_\rom{inner}$. 
    The \emph{solid} (\emph{dashed}) curves corresponds to the angular velocity of the $S^2$ ($S^1$) of the horizon. The  minimum inner radius is $r_\rom{m}\approx 0.95$.}
     \label{fig:SymSoln}
\end{figure}

{}Figure \ref{fig:SymSoln}(a) shows the dimensionless area $a_\rom{H}$ 
vs. the angular momentum $j^2$ (as introduced in \reef{jaH}) for the
symmetric bi-ring system (black curve).  
As suggested by the model in the previous section, the angular
momentum $j$ of the symmetric bi-ring system is unbounded from
above. When $\ka_1 \to 0$, $j\to \infty$ and the area goes to zero as
$a_\rom{H} \sim (4\sqrt{2}\, j)^{-1}$, in exact agreement with the non-interacting 
model of the previous subsection. In this limit the model gives a good description of the system, since the rings are long and thin,  hence far apart and with
negligible interactions. 

Just like the singly spinning black ring, there are two branches, a
thin and a fat ring branch, and they meet at a cusp, where the area
reaches its maximum and the angular momentum its minimum. At the cusp\footnote{The cusp is located at  
$\ka_1 = 
  \frac{1}{6} 
  \Big(
    1 - 2^{4/3} \big(5+\sqrt{57}\big)^{-1/3}
    + 2^{-1/3} \big(5+\sqrt{57}\big)^{1/3}
  \Big) 
  \approx 0.293$.}
the angular momentum and area take values
$j_\rom{min}^2 \approx 0.246$ and
$a_\rom{H\,max} \approx 0.485$.
The naive estimates of the non-interacting model in the previous subsection are different from the actual bi-ring results. We ascribe this to interactions between the two rings.

Each ring carries angular momentum in the plane of its $S^1$, and it does
not have intrinsic angular momentum on the $S^2$ (the
Komar angular momentum integral for the $S^2$ rotation vanishes). 
The 2-spheres are nonetheless rotating; they have non-vanishing angular velocity due to the interaction between the two rings. Through rotational frame-dragging, the $S^1$ rotation $J^{(1)}_\psi$ of ring 1 drags the $S^2$ of ring 2 to rotate with angular velocity $\Omega^{(2)}_\psi$. Ring 2 of course acts symmetrically on ring 1.
Claiming that the effect is gravitational frame-dragging is of course an interpretation of the physical properties of the solution, but it is very natural. The interpretation of similar dragging effects was tested in detail for the black saturn solution \cite{EF}. 

{}Figure \ref{fig:SymSoln}(b) shows the $S^1$ angular velocity 
$\omega^{(1)}_\psi$  for ring 1 and the $S^2$ angular velocity
$\omega^{(2)}_\psi$ for ring 2 plotted vs.~the 
inner horizon radius $r_\rom{inner}$ of ring 1. 
(This is done for the symmetric bi-ring configuration, so there is really no
distinction between parameters for ring 1 and ring 2.) 
The dimensionless angular velocity is defined as
\bea
  \omega_i &=& \sqrt{\frac{8}{3\pi}}\, \Omega_i\, (G \, M)^{1/2} \, .
\eea
The inner horizon radius is the $S^1$ radius on the
inside rim of the ring. Normalized by the total mass it is 
\beqa\label{rinner}
  r_\rom{inner} &=&
  (GM)^{-1/2}\sqrt{G_{\psi\psi}}\mid_{\rho=0,\bar{z}=\kappa2} 
  ~=~\sqrt{\frac{2(1-\ka_1-\ka_2)}{3\pi(\ka_2-\ka_1)}} \, .
\eeqa
The inner radius $r_\rom{inner}$ decreases monotonically along the symmetric ring branch in figure \ref{fig:SymSoln}(a). As $j \to \infty $, $r_\rom{inner} \to \infty$.

We use the inner horizon radius as a rough estimate of the spatial
separation between the two black rings. As shown in figure
\ref{fig:SymSoln}(b) the $S^1$ angular velocity $\omega_{S^1}$
increases monototically as the ring shrinks. As
$r_\rom{inner}$ decreases, the $S^2$ spins faster and faster,
reaching a maximum at $j^2=1/4$, where
$\omega_{S^2}=\omega_{S^1}=1$. Oppositely, as the separation
between the ring becomes large, $r_\rom{inner} \to \infty$ ($j\to
\infty)$, one finds that $\omega_{S^2} \to 0$. We also note that $\omega_{S^2}  < \omega_{S^1}$. These observations are
consistent with the interpretation that the $S^2$ spin is due to
gravitational frame-dragging.

In figure \ref{fig:SymSoln}(a) we also display the only other 
currently known vacuum black hole solution in 4+1 dimensions with \emph{equal}
magnitude angular momenta in the two planes of rotation: the
Myers-Perry black hole with $J_\psi = J_\phi$ (gray curve in figure \ref{fig:SymSoln}(a)). A few relevant
properties of the Myers-Perry black hole are reviewed in appendix \ref{app:MP}. 
At $j=0$ this is just the 4+1-dimensional Schwarzschild black
hole whose area is $a_\rom{H} = 2\sqrt{2}$. As $j$ increases, the area
decreases, and the curve ends at finite area.\footnote{Contrary to the singly
spinning Myers-Perry black hole which ends at $j=1$ and $a_\rom{H}=0$,
where it is singular.} The endpoint solution is a maximally rotating
extremal zero temperature black hole with 
$j^2 \equiv j_\phi^2=j_\psi^2=1/4$ and
$a_\rom{H} = \sqrt{2}$. 

The symmetric bi-ring branch in figure \ref{fig:SymSoln}(a) also has an endpoint at 
$j^2 \equiv j_\phi^2=j_\psi^2=1/4$, and indeed in this limit,\footnote{The limit is $\ka_1 \to 1/2$, corresponding to a collapse of
\emph{all} the finite length rods in the phase diagram. Due to the
balance condition, this particular scaling limit results in a non-trivial 
solution. The extremal limits of the Kerr black hole and the 4+1d Myers-Perry  black hole also  correspond to a controlled collapses of the horizon rods. This seems
  to be a general feature of zero temperature black holes.}  the temperature goes to zero. It is tempting to interpret the endpoint of the symmetric bi-ring
branch as an extremal zero temperature bi-ring configuration. However, examining a similar limit of the doubly spinning black ring in the next section, we are instead lead to the conclusion that the $j^2 \to 1/4$ limit of the symmetric bi-ring corresponds to a collapse of the rings to the symmetric zero temperature Myers-Perry black hole. In such a collapse, the horizon area will change discontinuously. Thus the $j^2=1/4$ endpoint of the symmetric bi-ring curve in figure \ref{fig:SymSoln}(a) is not part of the bi-ring phase.

Naturally one expects the existence of more general bi-ring systems in which each ring carries intrinsic angular momentum in both planes of rotation. Such solutions will extend the class of symmetric bi-rings, and their phases must also be considered for the full structure of the phase diagram figure \ref{fig:SymSoln}(a). Placing a black hole at the center of such a generalized bi-ring system will give rise to an even larger class of solutions, including symmetric ones.
Further discussion of generalizations follows in section \ref{s:disc}.


\section{Doubly spinning black rings}
\label{s:doubly}

The solution for the \emph{balanced} doubly spinning black ring was
presented by Pomeransky and Sen'kov \cite{PS}. We provide here an 
analysis of the doubly spinning ring solution and its physical properties.

\subsection{Analysis}
We use the solution\footnote{We
  have analytically verified that the solution presented in \cite{PS}
  indeed satisfies the Einstein vacuum equations, $R_{\mu\nu}=0$.}
in the form presented in \cite{PS}
  except that we interchange $\phi$ and $\psi$, so that $\phi$ is the
  azimuthal angle of the $S^2$ and $\psi$ parameterizes the circle of
  the ring.

The solution \cite{PS} is given in ring coordinates $(x,y)$ with
$|x|\le 1$ and $y\le-1$, and is parametrized by a scale $k$ and two
dimensionless parameters $\lambda$ and $\nu$ which are required to satisfy  
 \bea
   \label{range}
   0\leq\nu<1\, ,~~~~~~ 2\sqrt{\nu}\leq\lambda<1+\nu\, .
 \eea  

The balanced black ring \cite{ER} with rotation only in the plane of
the ring is found in the limit $\nu \to 0$. The unbalanced black ring
with angular momentum only on the $S^2$
\cite{Mishima:2005id,Tomizawa:2005wv,PF} cannot be obtained from the
Pomeransky-Sen'kov solution because the balance condition is imposed in
the solution presented in \cite{PS}. Likewise it is not possible to
obtain the full general 5-dimensional Myers-Perry black hole as a
``collapse'' limit of the balanced ring solution. The more general
unbalanced doubly spinning black ring metric  of Morisawa et al \cite{Morisawa:2007di} contains these limits.  

\vspace{4mm}
\noindent \emph{Asymptotics}:\\
Asymptotic coordinates $(\rho,\theta)$ are introduced through the
coordinate transformation
\bea
  x = -1 + 4 k^2 \, \alpha^2 \, \frac{1}{\rho} \cos^2\theta  \, ,
   \hspace{1cm}
  y = -1 - 4 k^2 \, \alpha^2  \, \frac{1}{\rho} \sin^2\theta \, ,
   \hspace{1cm}
  \alpha = \sqrt{\frac{1+\nu - \lambda}{1-\lambda}} \, .
\eea
In the $\rho \to \infty$ limit this brings the metric of \cite{PS} to
a manifestly flat form with the angular coordinates $\phi,\psi$ and
  the time coordinate $t$ canonically normalized. 

\vspace{4mm}
\noindent \emph{Horizon}:\\
The roots of the equation 
\bea
\label{horizon}
  1+\lambda \, y+\nu \, y^2 = 0
\eea
determine the locations of the inner and outer horizons;
the event horizon is located at
\beq
  y_h=\frac{-\lambda+\sqrt{\lambda^2-4\nu}}{2\nu} \, .
\eeq

The metric has a coordinate singularity at the roots of \reef{horizon}
where $g_{yy}$ diverges.
Good coordinates through the event horizon can be constructed by setting 
$\bar{y}=y - y_h$ and performing the coordinate transformation 
 \bea
   \label{ctransf}
   d\bar{\phi} = d\phi - \frac{A}{\bar{y}} \, d\bar{y} \, ,~~~~~
   d\bar{\psi} = d\psi - \frac{B}{\bar{y}} \, d\bar{y} \, ,~~~~~
   d\bar{t} = dt - \frac{C}{\bar{y}} \, d\bar{y} \, .
 \eea
The constants $A$, $B$, and $C$ are determined by eliminating potential
divergences in the metric components in the limit $\bar{y} \to 0$.  
In more detail: the constants $A$, $C$ are fixed in terms of $B$ (and
   $\lambda, \nu$) by requiring the absence of a $1/\bar{y}$  
divergence in $g_{t\bar{y}}$. This also eliminates similar potential
divergences in $g_{\phi \bar{y}}$ and $g_{\psi \bar{y}}$. Next $B$
can be chosen such that the divergences in $g_{\bar{y}\bar{y}}$
cancel. In the coordinates $(\bar{t},\bar{\phi},\bar{\psi},x,\bar{y})$,
the metric is then analytic across $\bar{y}=0$, and $y=y_h$ is
therefore the location of a regular event horizon. The analysis is
   valid for $2\nu^{1/2} < \lambda < 1+\nu$. 

An \emph{extremal} rotating black ring is obtained for
$\lambda=2\sqrt{\nu}$, when the inner and outer horizons coincide
(i.e.~the roots of \reef{horizon} coincide). Good coordinates through
the horizon can also be found in this case, but it requires a separate
analysis where terms $1/\bar{y}$ and $1/\bar{y}^2$ are both included
in the coordinate transformations \reef{ctransf}.  

\vspace{4mm}
\noindent \emph{Physical parameters}:\\
The ADM mass and angular momenta are
\bea
  &&M \; =\;   \frac{3 \pi\, k^2}{G} \frac{\lambda}{1+\nu-\lambda}\, ,
  \hspace{8mm}
  J_\phi\; =\;  
  \frac{4 \pi\, k^3}{G} 
  \frac{\lambda \, \sqrt{\nu \big[ (1+\nu)^2 - \lambda^2 \big]}}
         {(1+\nu-\lambda)(1-\nu)^2} \, , \\[2mm]
  &&J_\psi \; =\; 
  \frac{2 \pi\, k^3}{G} 
  \frac{\lambda \, (1+\lambda-6 \nu + \nu\, \lambda+\nu^2)\, \sqrt{ (1+\nu)^2 - \lambda^2}}
         {(1+\nu-\lambda)^2(1-\nu)^2} \, .
\eea
The angular velocities are
\bea
  &&\Omega_\psi \; =\;  \frac{1}{2 k} 
  \sqrt{\frac{1+\nu-\lambda}{1+\nu+\lambda}} \, ,
  \hspace{8mm}
  \Omega_\phi \; =\;  \frac{\lambda (1+\nu)-(1-\nu)\sqrt{\lambda^2 - 4\nu}}
    {4 k\, \lambda \sqrt{\nu}} 
    \sqrt{\frac{1+\nu-\lambda}{1+\nu+\lambda}} \, ,~~~~~~
\eea
and  the horizon area can be written
\bea
  A_\rom{H} &=& 
  \frac{32 \pi^2 k^3 \, \lambda (1+\nu+\lambda)}{(1-\nu)^2(y_h^{-1}-y_h)} \, .
\eea
The temperature can be found using the Smarr formula 
$\frac{2}{3} M = T_\rom{H} S + J_\phi \Omega_\phi + \Omega_\psi J_\psi$, with
$S=A_\rom{H}/(4G)$, and the result is
\bea
  T_\rom{H} &=& \frac{(y_h^{-1} - y_h) (1-\nu) \sqrt{\lambda^2 - 4 \nu}}{8\pi
    \, k\, \lambda (1+\nu +\lambda)} \, .
\eea
As expected, $T_\rom{H}=0$ for the extremal solution with $\lambda=2\sqrt{\nu}$.

Examining the ranges of the dimensionless angular
momenta (defined in \reef{jaH}) one finds
\bea
  j_\phi \le \frac{1}{4} \, ,\hspace{1cm}
  j_\psi \ge \frac{3}{4} \, ,
\eea
In particular the angular momenta can never be equal, and the ratio
$j_\phi/ j_\psi$ is less than or equal to $1/3$. 
The dimensionless angular velocities satisfy 
$0 < \omega_{\psi} \le 1$ and $0 \le \omega_{\phi} \le 1$.


\subsection{Phase diagram}
We are interested in three main questions: (1) how the $S^2$ angular
momentum $J_\phi$ changes the behavior of the spinning black ring,
(2) which regions of the phase diagram is covered by the doubly
spinning black rings, and (3) what are the zero temperature extremal limits. 
The answers are summarized in the phase diagrams
in figure \ref{fig:PSring}. 
There are several curves of interest in the phase diagram, and we
shall discuss them in turn.

\subsubsection*{The extremal black ring limit $\lambda \to 2\nu^{1/2}$}
The extremal black ring with $\lambda = 2 \nu^{1/2}$ is regular and
has zero temperature. Physically it corresponds to the $S^2$ rotating maximally, i.e.~saturating the Kerr bound. In this limit, the angular momentum $j_\psi$ and the area $a_\rom{H}$ can be written directly as functions of $j_\phi$:
\bea
 j_\psi ~=~ \frac{1+ 8 j_\phi^2}{8 j_\phi} \, ,~~~~~
 a_\rom{H} ~=~ 2\sqrt{2} \,  j_\phi \, .
\eea

\begin{figure}[t]
\begin{picture}(0,0)(0,0)
\put(-1,48){\small $a_\rom{H}$}
\put(77,-1){\small $j_\psi^2$}
\put(-3,42){\small $1$}
\put(-5,29){\small $\frac{1}{\sqrt{2}}$}
\put(5,-5){\small $\frac{9}{16}$}
\put(12,-5){\small $\frac{16}{25}$}
\put(46,-4){\small $1$}
\put(63,-4){\small $1.2$}
\put(84,48){\small $a_\rom{H}$}
\put(160,-1){\small $j_\psi^2$}
\put(82,42){\small $1$}
\put(80,29){\small $\frac{1}{\sqrt{2}}$}
\put(-5,29){\small $\frac{1}{\sqrt{2}}$}
\put(89,-5){\small $\frac{9}{16}$}
\put(97,-5){\small $\frac{16}{25}$}
\put(130,-4){\small $1$}
\put(147,-4){\small $1.2$}
\end{picture}
    \centering
        \includegraphics[height=4.5cm]{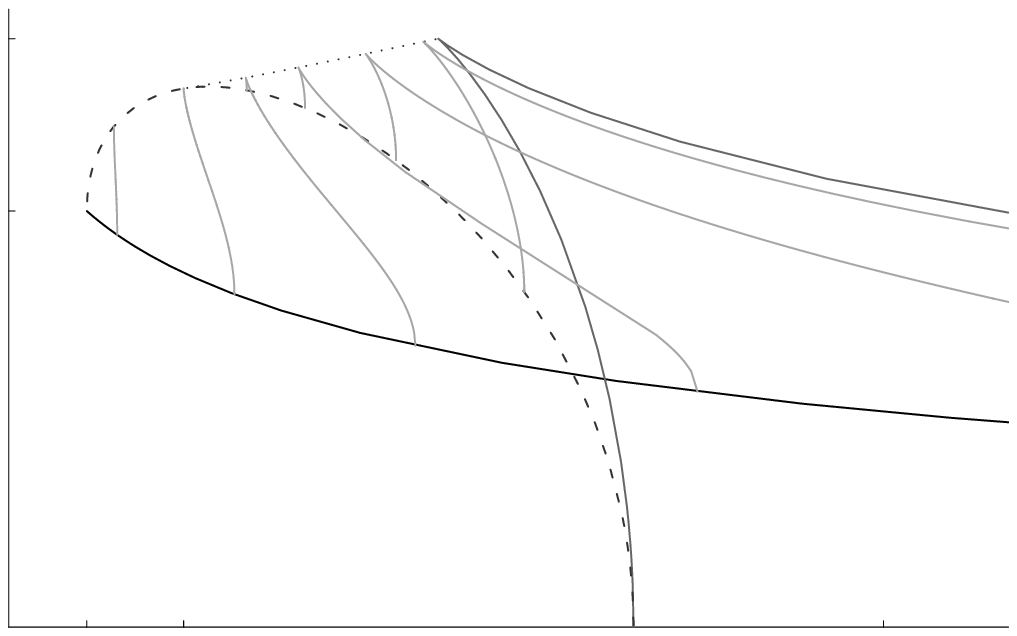}\hspace{1cm}
          \includegraphics[height=4.5cm]{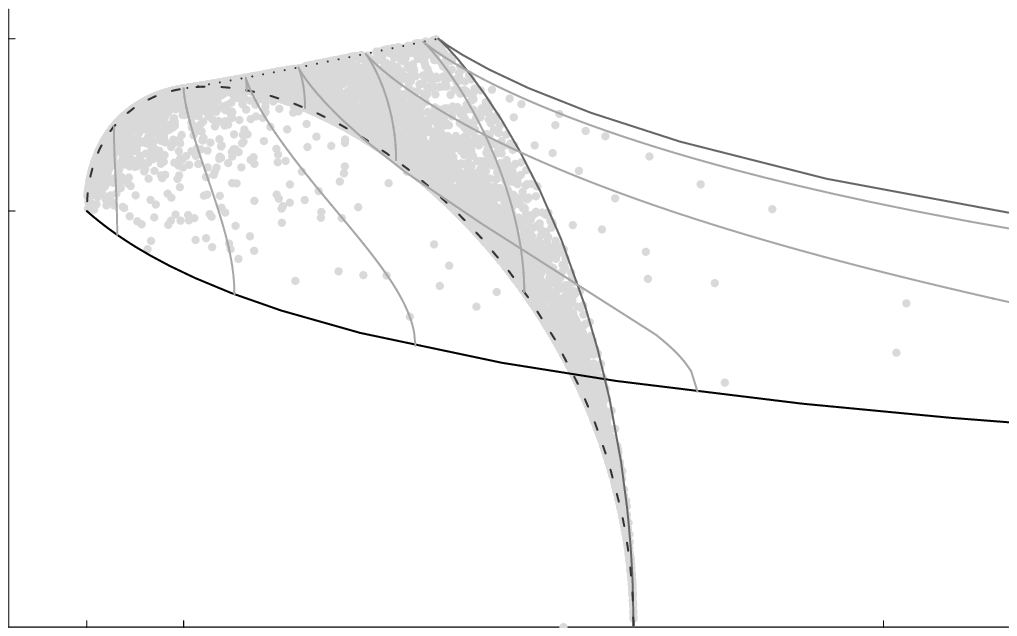}
        \bigskip
\caption{\small Phase diagrams for the doubly-spinning black ring: area
  vs.~the $S^1$ angular momentum $j_\psi^2$. The \emph{dark gray}
  curve shows the phase of the singly spinning black ring. The
  \emph{light gray} curves are branches of constant $j_\phi$; the
  particular values shown are (right towards left)  
$j_\phi^2=\frac{1}{500}, \frac{1}{100}, \frac{1}{50}, \frac{1}{35},
  \frac{1}{25}, \frac{1}{18}$. As $j_\phi^2 \to \frac{1}{16}$, its
  maximum value, the constant-$j_\phi$ curves degenerate to the point
  $j_\psi^2 = 9/16$ and $a_\rom{H} = 1/\sqrt{2}$. 
Each of the light gray constant-$j_\phi$ curves start at the
  zero-temperature branch of doubly spinning black ring solutions,
  shown as the \emph{solid black} curve, and they limit to the
  \emph{black dashed} curve (see main text). The \emph{black dotted} curve outlines the
  position of the cusp of the constant-$j_\phi$ curves; these are
  characterized by having minimum $j_\psi$ and maximum area for
  given fixed $j_\phi$. The plot on the right shows the same set
  of curves overlaying a scatter plot (\emph{lighest gray} --- 10,000
  points) indicating the range in the phase diagram covered by doubly
  spinning black rings. The region is bounded by the
  curves we have described.} 
    \label{fig:PSring}
\end{figure}

In figure \ref{fig:PSring} this extremal phase is shown as the solid
black curve. It starts at $j_\psi = 3/4$, $j_\phi =1/4$ and $a_\rom{H}=1/\sqrt{2}$,
and the area decreases monotonically to zero as $j_\psi \to
\infty$ ($j_\phi \to 0$). We conclude that there exists zero temperature black rings
for any $S^1$ angular momentum $j_\psi > 3/4$. 

In the extremal ring limit, $\lambda=2\sqrt{\nu}$, the inner and outer horizon radii are the same $r^\rom{(inner)}=r^\rom{(out)}$. This is similar to the case of supersymmetric black rings \cite{EEMR1,EEMR2,BW,GG}.

\subsubsection*{The second extremal limit $\nu \to 1, \lambda \to 2$ is a collapse}
A second extremal limit corresponds to the limit
$\nu \to 1$. Due to the parameter restrictions \reef{range}, we have
to take  $\lambda \to 2$ as $\nu \to 1$. This must be done in a way
that leaves dimensionless parameters finite, and that requires the ratio  
\bea
  \sigma = \frac{1+\nu -\lambda}{(1-\nu)^2}
\eea
to be finite in the limit. Solving this equation for $\lambda$ and taking the limit $\nu \to 1^-$ gives 
\bea
  T_\rom{H} = 0\, ,~~~~~
  j_\phi = \sigma \, ,~~~~~
  j_\psi = 1-\sigma \, ,~~~~~
  a_\rom{H} = \frac{4\sqrt{2}\,\sigma^{3/2}}{1-\sqrt{1-4\sigma}} \, ,~~~~~
  \omega_{\psi,\phi} \to 1 \, .
\eea
The $(j_\psi^2,a_\rom{H})$ curve is shown in figure \ref{fig:PSring} (dashed curve).
As we now show, this curve does \emph{not} represent a phase of extremal rings.

The limit $\nu \to 1, \lambda \to 2$ appears singular, but this is just a coordinate artifact, and the resulting solution is actually the extremal Myers-Perry black hole with parameters $a_1$, $a_2$ and $\mu^{1/2} = a_1 + a_2$. To see this, define
\bea
   \mu ~=~ \frac{k^2}{(1-\nu)^2\, \sigma} \, ,~~~~
   a_1 ~=~ \mu^{1/2} (1-\sigma) \, ,~~~~
   a_2 ~=~ \mu^{1/2} \, \sigma \, ,
\eea
and perform the following coordinate transformation
\bea
  x= -1 + \frac{16 \sqrt{a_2}\, k^3 \cos^2{\theta}}{(a_1+a_2)^{3/2}(r^2 - a_1 a_2)}
  \, ,~~~~~
  y= -1 - \frac{16 \sqrt{a_2}\, k^3 \sin^2{\theta}}{(a_1+a_2)^{3/2}(r^2 - a_1 a_2)} \, .
\eea
When $\nu \to 1$ one finds the metric for the extremal Myers-Perry solution.

In this collapse limit of the ring to an extremal Myers-Perry black hole, the area is discontinuous, just like in the similar collapse limits of supersymmetric black rings \cite{EEMR2}. So the dashed curve in figure  \ref{fig:PSring}  does not itself represent a phase of black rings, but it marks the endpoint of the fixed-$j_\phi$ phases which are described below; at the endpoint the ring has collapsed to the zero temperature Myers-Perry black hole.\footnote{We had examined the inner and other horizon radii and observed that they remain finite in the limit $\nu \to 1, \lambda \to 2$. This may seem to contradict the collapse of the ring, but one must keep in mind that these radii are defined by the horizon geometry which is discontinuous in the limit.}

\subsubsection*{Curves of constant $S^2$ angular momentum}
The dimensionless angular momentum on the $S^2$ is given by 
\bea
  j_\phi^2 = \frac{\nu
    (1+\nu-\lambda)^2(1+\nu+\lambda)}{2\lambda(1-\nu)^4} \, . 
\eea
Finding the fixed-$j_\phi$ curves in the phase diagram can be done by
solving $j_\phi=j_\phi^*$ for some fixed value
$0 \le j_\phi^* \le 1/4$. It is a cubic equation in $\lambda$ which
for $0<\nu<1$  and $0< j_\phi < 1/4$ has three real 
roots. Two of these roots always violate the constraints \reef{range},
but the third root satisfies \reef{range} provided $\nu >
\nu_c(j_\phi^*)$, where $\nu_c(j_\phi^*)$ is some critical value which
depends on the value $j_\phi^*$. 

Figure \ref{fig:PSring} shows (in light gray) curves of fixed 
$j_\phi^2=\frac{1}{500}, \frac{1}{100}, \frac{1}{50}, \frac{1}{35},
  \frac{1}{25}, \frac{1}{18}$. For small $j_\phi$ the curve is very
  similar to the singly spinning black ring with $j_\phi =0$, shown in dark gray.
The fixed-$j_\phi$ curves have two endpoints at non-vanishing area $a_\rom{H}$. One
corresponds to the limit $\nu \to \nu_c(j_\phi^*)$ which is  
equivalent to $\lambda \to 2\nu^{1/2}$; this is the zero temperature
extremal branch (solid black curve). The other endpoint of the fixed-$j_\phi$ curves is at $\nu \to 1$, and as shown above the ring collapses to an extremal Myers-Perry black hole in this limit.

When $j_\phi < 1/5$ the fixed $j_\phi$-curves have two branches: a
thin and a fat ring branch, just like the singly spinning black
ring. When  $1/5 \le j_\phi \le 1/4$ only the thin ring branch exists.

\subsubsection*{Curve of cusps}
The dotted black curve in figure \ref{fig:PSring} are the position of
the cusp of the constant $j_\phi$ curve, where $j_\psi$ is minimized
and the area maximized. The cusp curve is found by using a Lagrange
multiplier to fix $j_\phi$ while extremizing $j_\psi$. This fixes
$\lambda$ in terms of $\nu$ as 
\bea
  \lambda = \frac{1}{4} \Big( -1 -\nu + \sqrt{(9 + \nu)(1+9 \nu)}\Big) \, . 
\eea
The cusp only exists for sufficiently small $j_\phi$, namely $j_\phi <
1/5$. The fixed $j_\phi = 1/5$ curve ends at $j_\psi = 4/5$ and
$a_\rom{H}=\frac{2}{5}\sqrt{3+\sqrt{5}}$, where the dotted curve of
cusps ends on the dashed curve.

\subsection{Physics discussion}

One feature of the curves of fixed angular momentum $j_\phi$ on the
$S^2$ is that the $S^1$ angular momentum cannot be arbitrarily
large. This is easily understood. Recall that in the limit of large
$j_\psi$, the singly rotating black ring becomes large and
thin. Non-zero $j_\phi$ means that the $S^2$ part of the ring behaves
like a Kerr black hole, in particular it will have to obey the Kerr
bound on the angular momentum. This implies that for given non-vanishing
$j_\phi$, the size of the $S^2$ cannot become arbitrarily thin (the
effective $S^2$ mass cannot be too small because that would violate
the Kerr bound). This in turn means that one cannot spin up the ring
to arbitrarily large $j_\psi$. Hence for given $j_\phi$, there is a
maximum possible value for $j_\psi$. When $j_\phi$ is small, the
maximum value is large, but decreases as $j_\phi$ increases, as is
seen in figure \ref{fig:PSring}.

Another qualitative feature is the disappearance of the ``fat ring
branch'' as $j_\phi$ becomes large. Consider two diagonally
opposite 2-spheres of the ring. They both carry $j_\phi$ angular
momentum which creates an attractive spin-spin interaction
\cite{wald}. As the (inner) $S^1$ decreases, the strength of this
spin-spin attraction grows stronger, and it becomes harder to balance
the ring by angular momentum on the $S^1$. This is what causes the
diminishing of the fat ring branch as $j_\phi$ increases, and for
$j_\phi \ge 1/5$, the fat branch disappears.\footnote{We are grateful to R.~Emparan for discussions of these properties.}

The analysis of \cite{EEV} suggested that the thin black ring branch
solutions are stable to radial perturbations and the fat rings
unstable. This was in concordance with an expected result
\cite{Arcioni} that at least one extra mode of instability appeared when going
across the cusp from the thin black ring branch to the fat black ring branch.  
Extrapolating these results, doubly spinning rings with
large enough $S^2$ angular momentum, $j_\phi \ge 1/5$, may be expected to be radially stable.


\section{Zero-temperature bicycles}
\label{s:TzBike}

We found in section \ref{s:sym} that the symmetric bi-ring system had
a zero-temperature limit. In the previous section we investigated a similar 
zero temperature limit of the doubly spinning black ring, and showed that it corresponded to the collapse of the ring to an extremal Myers-Perry black hole.
In this section we take a scaling limit to find a similar collapse zero-temperature limit of the bi-ring solution.

For the symmetric bi-rings the limit of zero temperature
is $\ka_1, \ka_2 \to \ka_3 = 1/2$. It is therefore natural to take 
the limit $\ka_i \to \ka$ for some $0<\ka <1$ for the bi-ring solution. This limit corresponds to collapsing all the finite rods in the rod diagram to zero 
length.

Here we want to take a limit such that $T_1,T_2 \to  0$, but the mass,
angular momenta, angular velocities, and horizon areas remain
finite. The simplest approach is to first take the limit $\ka_i \to
\ka$ and then impose the balance conditions, so we also need the RHS
of the balance 
conditions \reef{bal1}-\reef{bal2} to be finite in the limit. 
Without loss of generality we pick $\ka = \ka_1$, and the desired scaling limit is then obtained by setting 
\bea
  \label{scaling}
  \ka_2 = \ka_1 + w_1\, \eps\, , ~~~~~
  \ka_3 = \ka_1 + w_2\, \eps\, , ~~~~~
  \ka_4 = \ka_1 + w_3\, \eps\, , ~~~~~
  \ka_5 = \ka_1 + \eps\, , ~~~~~
\eea
and taking $\eps \to 0$. (By exploiting the freedom of rescaling
$\eps$, we have set the coefficient of $\eps$ in $\ka_5$ equal to 1.) The 
ordering of the $\ka_i$ implies
\bea
  0 < w_1 < w_2 < w_3 < 1 \, .
\eea
The balance conditions become
\bea
&&
1~=~\frac{w_3(1-w_1)(w_2-w_1)(w_3-w_1)}{w^2_2\,(1-\ka_1)}\, ,
\hspace{8mm}
1~=~\frac{w_3(1-w_1)(w_3-w_1)(w_3-w_2)}{(1-w_2)^2\,\ka_1}\,.~~~~~
\eea
One finds that as $\eps \to 0$, the temperatures go to zero linearly in $\eps$, so the
ratio of the temperatures is finite in the limit,
\bea
  \label{T1oT2}
  \xi = \lim_{\eps \to 0}\left( \frac{T_1}{T_2} \right) = 
  \frac{w_2}{(1-w_2)} \sqrt{\frac{(1- w_1)(1-w_3)}{w_1 \, w_3}} \, .
\eea
This ratio can take any non-negative value, and the limit
$T_1,T_2 \to 0$ can therefore be taken along any subfamily of solutions with $T_1 = \xi\, T_2$ for
any value of $\xi$. 

In the scaling limit we find (superscript ``$(0)$'' indicates zero temperature)
\bea
  &&M^{(0)} ~=~ \frac{3 \pi \, L^2}{4G} \, , \hspace{8mm}
  J^{(0)}_\phi ~=~
  \frac{\pi\, L^3}{\sqrt{2}\, G}\,  (1-\ka_1) \, ,\hspace{8mm}
  J^{(0)}_\psi ~=~
  \frac{\pi\, L^3}{\sqrt{2}\, G}\,  \ka_1 \, .
\eea
Note that the relationship between the mass-fixed angular momenta are exactly the same as for the extremal Myers-Perry black hole. We therefore expect the limit to be a collapse of the two orthogonal rings to a single extremal Myers-Perry black hole, just as in the $\nu \to 1, \lambda \to 2$ limit of the doubly spinning black ring. Again, the horizon areas will be discontinuous in this limit, so the corresponding limiting curves in the phase diagram do not correspond to physical black hole phases. 

Even if the zero temperature limit of the bi-ring solution presented in this paper turns out to simply give the extremal Myers-Perry black hole, we do expect there to exist zero temperature bi-ring system. These are obtained from a more general bi-ring system, constructed from two rings which each carry intrinsic angular momenta
on both $S^1$ and $S^2$. For this more general bi-ring the limit of
extremality is similar to the $\lambda \to 2 \sqrt{\nu}$ limit of each of the
doubly spinning black rings of the system. We expect those extremal solutions to have
arbitrarily large angular momenta in both planes. In the next section we discuss the zero temperature phase diagram.

\section{Discussion}
\label{s:disc}

In this paper we have analyzed two 4+1-dimensional
asymptotically flat black hole vacuum solutions: one is the doubly spinning black ring, the other  
is the bicycling black rings, the bi-rings,
 consisting of two black rings balanced in orthogonal
planes. Both solutions are constructed by the inverse scattering
method. The doubly spinning black ring was found by Pomeransky and
Sen'kov \cite{PS}, and the bi-ring solution constructed here is new. 

We showed in section \ref{s:doubly} that the doubly spinning black ring had two limits of zero temperature. The first of these limits, $\lambda \to 2\sqrt{\nu}$, gives a zero-temperature extremal doubly rotating black ring. 
We have proven that good coordinates exist across the horizon, so indeed
the solution is regular everywhere on and outside the
horizon. This branch of zero temperature vacuum solution shares
certain features with supersymmetric black rings, for instance that
the inner and outer horizon radii are the same. However, there are also
clear differences, for example that the extremal non-supersymmetric
rings have non-vanishing angular velocities, while $\Omega=0$ for
asymptotically flat supersymmetric black holes. 
This difference is significant for studying dragging in the system.
Remarkably, Reall has shown \cite{harvey} that the entropy of this non-supersymmetric extremal black ring can be reproduced from a microscopic calculation.

In the other extremal limit ($\nu \to 1$,$\lambda \to 2$) of the doubly spinning ring, we have shown that the ring collapses to a zero-temperature Myers-Perry black hole.

\begin{figure}[t]
\begin{picture}(0,0)
\put(2,44){\small $a_\rom{H}$}
\put(71,-1){\small $j_{\psi}$}
\put(63.6,-4){\footnotesize $1.5$}
\put(44,-4){\footnotesize $1$}
\put(33.3,-4){\footnotesize $\frac{3}{4}$}
\put(23,-4){\footnotesize $\frac{1}{2}$}
\put(-4,38){\footnotesize $\sqrt{2}$}
\put(-1,27){\footnotesize $1$}
\put(-3.5,19){\footnotesize $\frac{1}{\sqrt{2}}$}
\put(34,38){\footnotesize MP black hole}
\put(47,17.5){\footnotesize doubly spinning}
\put(47,13.5){\footnotesize black ring}
\put(82,44){\small $a_\rom{H}$}
\put(151.5,-1){\small $j_{\phi}$}
\put(146,-4){\footnotesize $1$}
\put(114.5,-4){\footnotesize $\frac{1}{2}$}
\put(98.5,-4){\footnotesize $\frac{1}{4}$}
\put(77,38){\footnotesize $\sqrt{2}$}
\put(80,27){\footnotesize $1$}
\put(78,19){\footnotesize $\frac{1}{\sqrt{2}}$}
\put(134,34){\footnotesize MP black hole}
\put(99,13.5){\footnotesize doubly spinning}
\put(99,9.5){\footnotesize black ring}
\put(32,-12.5){\footnotesize (a)}
\put(115,-12.5){\footnotesize (b)}
\end{picture}
\vspace{8mm}
    \centering{
        \includegraphics[width=6.5cm]{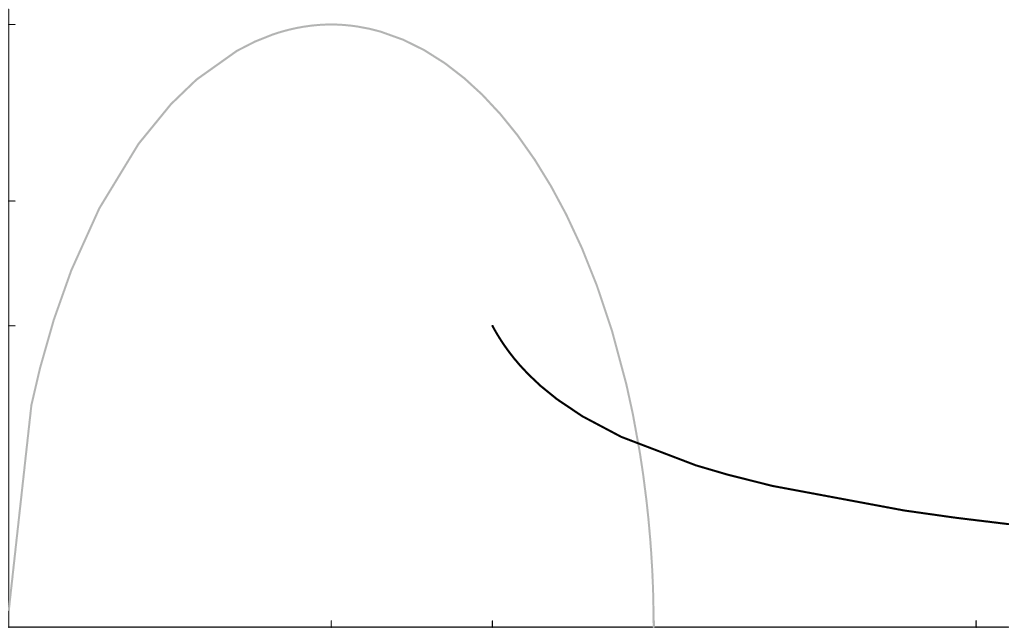}
        \hspace{1.3cm}
        \includegraphics[width=6.5cm]{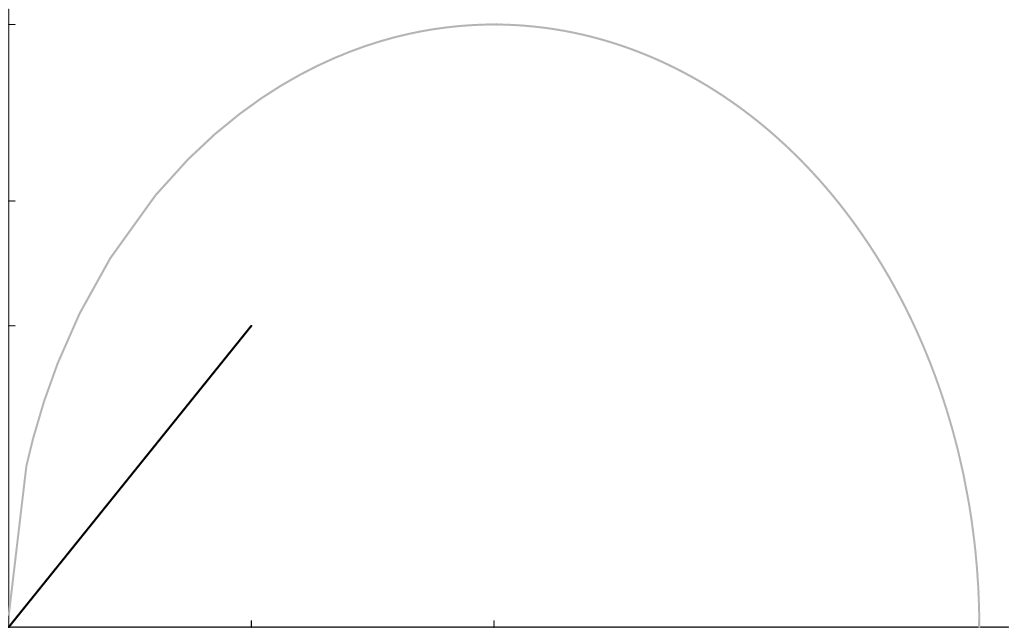}}
    \vspace{3mm}
    \caption{\small Phase diagram for zero temperature single horizon
    black holes. In figure (a) we choose the $S^1$ angular momentum of
    the doubly spinning ring on the horizontal axis, in figure (b) it
    is the $S^2$ angular momentum.  
    The \emph{gray} curve is the zero temperature Myers-Perry black
    hole. Its entropy is maximized when the system is symmetric,
    i.e.~when $j_\psi = j_\phi = 1/2$. 
    The \emph{solid} black curve is the
    regular zero temperature 
    limit $\lambda \to 2 \nu^{1/2}$ of the doubly spinning black
    ring. It reaches maximum area $a_\rom{H}=1/\sqrt{2}$ for maximal
    $j_\phi=1/4$ and minimal $j_\psi = 3/4$.}
     \label{fig:zeroTbhs}
\end{figure}

Figure \ref{fig:zeroTbhs} shows the known phases of zero temperature
single horizon vacuum black holes: the zero temperature Myers-Perry
black hole and the extremal limit of the doubly spinning black
ring. The two phase diagrams, (a) and (b), shown in figure
\ref{fig:zeroTbhs} are equivalent. In figure (a) we use the $S^1$
angular momentum $j_\psi$ of the ring as the ``order parameter'' on
the horizontal axis, and in figure (b) it is the $S^2$ angular
momentum. There is no non-uniqueness in the phase diagram in figure \ref{fig:zeroTbhs}, because the two solutions never co-exist with the same $j_\psi$ and $j_\phi$.

In addition to the single horizon phases shown in figure
\ref{fig:zeroTbhs} there will exist other zero temperature phases for
4+1-dimensional vacuum black holes. These include extremal bi-rings and extremal saturn solutions.
The bi-rings we have constructed in this paper are a subclass of a more
general family of bi-ring solutions. Recall that our bi-rings are
essentially superpositions of singly spinning black rings. Obviously
one can construct more general solutions by superimposing the
doubly spinning black rings. This would presumably require using a
two-step solution generating procedure with the second step involving 
a non-diagonal seed, as in  \cite{PS}.

When the $S^2$'s of the generalized bi-ring system carry maximal spin, one expects to have a zero temperature bi-ring configuration consisting of two doubly spinning extremal black rings. The general bi-ring system is expected to have a 3-fold continuous non-uniqueness corresponding to the freedom of distributing the mass and the two angular momenta between the two black objects while keeping the total asymptotic ADM mass and angular momenta fixed and imposing balance. Requiring zero temperature for each ring gives two constraints, and the degeneracy reduces to 1-fold continuous non-uniqueness for the extremal bi-rings.  Thus these solutions will fill up a 2d area of the phase diagram.

\begin{figure}[t]
    \centering{
\includegraphics[width=7.5cm]{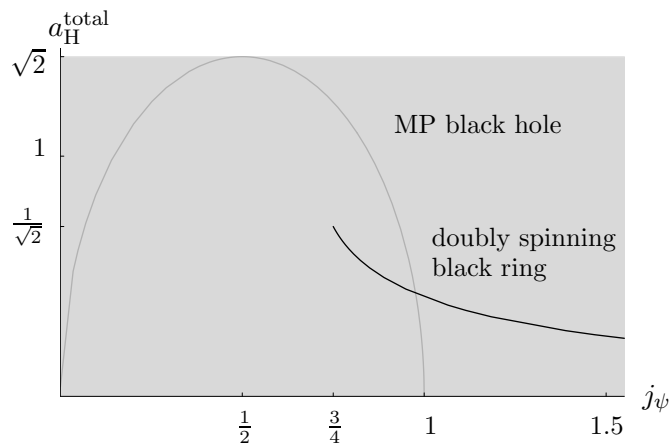}}
\begin{picture}(0,0)
\put(-78,48){\small $a_\rom{H}^\rom{total}$}
\put(1,-1){\small $j_{\psi}$}
\put(-6,-5){\footnotesize $1.5$}
\put(-28,-5){\footnotesize $1$}
\put(-41,-5){\footnotesize $\frac{3}{4}$}
\put(-53,-5){\footnotesize $\frac{1}{2}$}
\put(-83,43){\footnotesize $\sqrt{2}$}
\put(-80,32){\footnotesize $1$}
\put(-83,22){\footnotesize $\frac{1}{\sqrt{2}}$}
\put(-32,35){\footnotesize MP black hole}
\put(-27,20){\footnotesize doubly spinning}
\put(-27,16){\footnotesize black ring}
\end{picture}
    \vspace{3mm}
    \caption{\small Expected phase diagram for zero temperature
    extremal 4+1-dimensional asymptotically flat vacuum black
    holes. Extremal bi-ring saturn configurations are
    anticipated to fully cover the gray strip with $0 \le j < \infty$
    and $0<a_\rom{total} < \sqrt{2}$. In addition, generalized extremal bi-ring solutions will provide further non-uniqueness in the system.} 
     \label{fig:Tzphases}
\end{figure}

And even more non-uniqueness can be expected. 
Figure \ref{fig:Tzphases} shows our expectations for the full phase
diagram of \emph{extremal} doubly rotating 4+1-dimensional
asymptotically flat vacuum black holes. The curves are the single
horizon solutions shown also in figure \ref{fig:zeroTbhs}(a). The gray
strip covers the phase diagram from $j=0$ to arbitrarily large $j$. The total area is bounded
by the maximum of the zero temperature Myers-Perry black hole, so that the gray
strip covers $j \ge 0$ and $0<a_\rom{H}^\rom{total} <\sqrt{2}$. The proposal is that there exist zero temperature black hole solutions at any point of the gray strip.

The analysis behind this is similar to the one done for the phase diagram
of the non-extremal singly spinning black holes in \cite{EEF}. One way
to justify that the whole strip in the phase diagram is covered by
solutions is to consider the system of a zero temperature Myers-Perry
black hole surrounded by a zero temperature doubly spinning black
ring; this is the natural zero temperature black saturn
solution. One can get arbitrarily close to the upper bound
$a_\rom{H}^\rom{total}=\sqrt{2}$ by putting most of the mass in the
Myers-Perry black hole, and tuning its angular momentum to maximize
its entropy. 
The \emph{total} angular momentum in one plane of rotation can 
then be adjusted to be any value $0 \le j < \infty$ by
including a thin zero temperature black ring around the black
hole. The angular momentum in the orthogonal plane can likewise be
adjusted to any value by including a second black ring, thus combining
the bi-rings and black saturn into a black bi-ring saturn.\footnote{Having arbitrarily
large angular momentum relies on the space being infinite since the
ring needs to be large. So if the system is put in a box, the phase
structure would necessarily change.}

We expect that any
balanced zero temperature black hole configuration in 4+1-dimensional
vacuum gravity, with asymptotically flat boundary conditions, have
total dimensionless area bounded by $a_\rom{H}^\rom{total}=\sqrt{2}$,
which is the maximal area of the zero temperature Myers-Perry black
hole. It is achieved for $j_\psi = j_\phi$. For comparison, it was argued in \cite{EEF} that non-extremal
black hole configurations exist of any $j$ and any total area
$a_\rom{H}^\rom{total}\le 2\sqrt{2}$. The inequality is saturated only for
the static 4+1-dimensional Schwarzschild solution.

The continuous non-uniqueness in the general non-extremal black hole
phase diagram can be arbitrarily large. A system consisting of $n$
concentric doubly spinning black rings distributed at will in the two
orthogonal planes and surrounding a central doubly spinning black hole
will have $3n$-fold continuous non-uniqueness. There are only 3
conserved quantities, the ADM mass and two angular momenta, but these
can be distributed continuously (at least classically) between the
$n+1$ objects (subject to balance conditions). In addition to
this, there can be discrete non-uniqueness.
Imposing thermodynamic equilibrium, equal temperatures and equal
angular velocities, is expected to fix the radius of the rings, so the
most general equilibrium system would be the bi-ring saturn discussed above.

Higher dimensional gravity has proven to contain an intriguing richness of black hole solutions. We are privileged to be able to access some of them as exact solutions. Although these are in some cases rather involved solutions, like black saturn and the bi-rings, it is possible to extract interesting physics. In general, we cannot expect to be lucky to have exact solutions for all interesting black hole configurations, and other methods for constructing solutions are needed. Recently the matched asymptotic expansion method was used to construct large radius higher dimensional black rings \cite{Emparan:2007wm}. From these solutions interesting properties of the phase diagram of black holes were extracted and the remaining structure of the phase diagram conjectured. Just like black saturn had a place in that phase diagram, so will doubly spinning black rings, bi-rings and bi-ring saturns in the generalization to involve spin in both planes of rotation.

In this paper we have examined extremal limits of exact black hole solutions. A common picture seems to be that if a rod diagram description of the solution can be given, then the zero temperature extremal limit is a scaling limit in which the finite length rods shrink to zero size. This characterization may lead to simplifications for constructing extremal black hole configurations; it would be interesting and useful if it could be implemented directly in the inverse scattering method.


\section*{Acknowledgments}

We are grateful to Roberto Emparan for very useful discussions and for
collaboration during the early stages of this work. We are also grateful to Harvey Reall and Pau Figueras for their help with clarifying the second extremal limit of the doubly spinning black ring.

We would like to thank Joan Camps, Troels Harmark, Gary Horowitz, Jim
Liu, Niels Obers, Don Marolf, Oscar Varela and Amitabh Virmani for useful discussions.  
Veronika Hubeny and Mukund Rangamani deserve special thanks for
numerous lively discussions and for sharing with us their ideas for higher dimensional  black hole configurations.

We thank the organizers of the ÒPre-Strings 2007Ó workshop held in Granada, June 2007, where work on the bicycling ring solutions was presented. MJR would like to thank the organizers of the ÒSimons Workshop 2007Ó, Stony Brook University, and the Physics Department at the University of California at Santa Barbara, where part of this work was done, for their hospitality. HE thanks CERN for hospitality during the final stage of this work. 

MJR is partially supported by the FI scholarship from Generalitat de Catalunya, the European Community FP6 program MRTN-CT-2004-005104 and MCYT FPA 2004-04582-C02-01.
HE is supported by a Pappalardo Fellowship in Physics at MIT and in part by the US Department of Energy through cooperative research agreement DE-FG0205ER41360

\appendix

\section{Horizon metrics}
\label{app:hori}

The BZ parameters introduced by the soliton transformations are dimensionful. It is more natural to rescale them $b_2 L^{-1} \rightarrow  b_2$ and $c_1 L^{-1} \rightarrow c_1$  to make them dimensionless. The rescaled parameters are used in this appendix. It is implicitly understood that $c_1$ and $b_2$ are fixed by \reef{thec1b2}.

The bi-ring solution has two disconnected horizons, each with topology $S^1\times S^2$. Here we give the metric on the event horizons. 
The ring with event horizon located at $\rho=0$ and $\kappa_1\le \bar{z} \le \kappa_2$ is refered to as ring 1, and the one with event horizon at $\rho=0$ and $\kappa_4\le \bar{z} \le \kappa_5$  is ring 2.


\subsubsection*{Black ring 1}
The metric on the horizon of ring 1 is
\bea
  \nonumber
  ds^2_\rom{BR1} &=&
  \frac{L^2\, s_\rom{BR1}^2 \, f_1(\bz)}
       {(\bz-\ka_1)(\ka_2-\bz)}  \, d\bz^2 \\[2mm]
  \nonumber
  &&+
  \frac{L^2\, \bz\, (\bz-\ka_1)(\ka_2-\bz)^2}
       {(1-\bz) (\ka_4 - \bz) \, f_1(\bz)}
  \left[
    b_2^{-1} d\psi
    + c_1^{-1}\, \frac{\ka_1\, (1- \bz)}{\bz} \, d\phi
  \right]^2 \\[2mm]
  &&
      + \frac{L^2\, (\bz-\ka_1)^2 (\ka_2-\bz)(\ka_4-\bz)}
         {2\, \bz (\ka_3-\bz)(\ka_5-\bz)\, f_1(\bz)}\, d\phi^2 \, 
      + \frac{L^2\,  (\ka_3 - \bz)(\ka_5 - \bz)}
        {2\, (1-\bz)\, f_1(\bz)}\, d\psi^2 \, ,
\eea
where $\bar{z} \in [\ka_1, \ka_2]$ and
\bea
  s_\rom{BR1} &=&
  \frac{\sqrt{2\, \ka_2\, \ka_3\, \ka_5\,
    (1-\ka_1)(\ka_2-\ka_1)(\ka_4-\ka_1)}}
       {\ka_4 \, (\ka_3-\ka_1)(\ka_5-\ka_1)} \, , \\[3mm]
  \nonumber
  f_1(\bz) &=&  \frac{(\bz-\ka_1)(\ka_4-\bz)}
    {4 \, \bz\, (1-\bz)}
    + c_1^{-2}\, \frac{\ka_1^2\, (\ka_2-\bz) (\ka_3-\bz) (\ka_5-\bz)}
    {2 \, \bz\, (\ka_4-\bz)} 
    + b_2^{-2} \, \frac{(\bz-\ka_1)^2(\ka_2-\bz)^2}
      {2 (1-\bz) (\ka_3-\bz)(\ka_5-\bz)} \, .
\eea
The horizon area is
$A_1 = (2\pi)^2\,  L^3 \, (\ka_2-\ka_1)\, s_\rom{BR1}$. 


\subsubsection*{Black ring 2}
The metric on the horizon of ring 2 is
\bea
  \nonumber
  ds^2_\rom{BR2} &=&
  \frac{L^2\, s_\rom{BR2}^2 \, f_2(\bz)}
       {(\bz-\ka_4)(\ka_5-\bz)}  \, d\bz^2 \\[2mm]
  \nonumber
  &&+
  \frac{L^2\,  (1-\bz)(\bz-\ka_4)^2(\ka_5-\bz)}
       {4\, \bz\, (\bz - \ka_2) \, f_2(\bz)}
  \left[
    c_1\, \ka_5^{-1}\, d\phi + b_2\,  \frac{(1-\ka_5)}{(1-\ka_1)}
            \frac{\bz }{(1-\bz)} \, d\psi
  \right]^2 \\[2mm]
  &&  + \frac{L^2\,  (\bz-\ka_1)(\bar{z}-\ka_3)}
        {2\, \bz\, f_2(\bz)}\, d\phi^2
      + \frac{L^2\, (\bz-\ka_2) (\bz-\ka_4)(\ka_5-\bz)^2}
         {2(\bz-\ka_1)(\bz-\ka_3)(1-\bz)\, f_2(\bz)}\, d\psi^2 \, ,
\eea
where $\bar{z} \in [\ka_4, \ka_5]$ and
\bea
  s_\rom{BR2} &=&
  \frac{\sqrt{2\, \ka_5\, (1-\ka_1)(1-\ka_3)(1-\ka_4)
       (\ka_5-\ka_4)(\ka_5-\ka_2)}}
       {(1-\ka_2)(\ka_5-\ka_1)(\ka_5-\ka_3)} \, , \\[3mm]
  \nonumber
  f_2(\bz) &=&  \frac{(\bz-\ka_2)(\ka_5-\bz)}
    {4 \, \bz\, (1-\bz)}
    + c_1^2\, \frac{(\bz-\ka_4)^2(\ka_5-\bz)^2}
    {8\, \ka_5^2 \, \bz\, (\bz-\ka_1)(\bz-\ka_3)} \\[1mm]
    \nonumber
    &&+ b_2^2 \frac{(1-\ka_5)^2(\bz-\ka_1)
            (\bz-\ka_3)(\bz-\ka_4)}
      {8 (1-\ka_1)^2 (1-\bz) (\bz-\ka_2)} \, .
\eea
The horizon area is $A_2 = (2\pi)^2\,  L^3 \, (\ka_5-\ka_4)\, s_\rom{BR2}$.

\vspace{3mm}
Note that the functions $f_1$ and $f_2$ are manifestly positive.


\section{Myers-Perry Black Hole}
\label{app:MP}

\begin{figure}
\begin{picture}(0,0)
\put(1,57){\small $a_\rom{H}$}
\put(88,3){\small $j_{\psi}^2$}
\put(42,-1){\small $0.5$}
\put(83,-1){\small $1$}
\put(-2,20){\small $1$}
\put(-2,37){\small $2$}
\end{picture}
    \centering
        \includegraphics{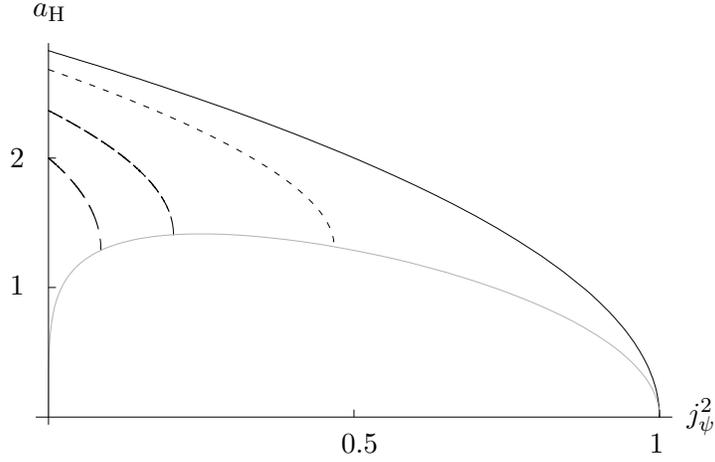}
    \caption{\small Phase diagram for the Myers-Perry black hole in five dimensions. The  \emph{dashed} lines show to Myers-Perry black hole phases for fixed values of $j^2_{\phi}=\,1/10,\,3/10,\,5/10$ (right to left). The five dimensional singly spinning black hole, \emph{solid} curve, becomes singular in the limit $j_\psi \to 1$ The \emph{gray} curve is the phase of the zero temperature extremal Myers-Perry black holes.}
     \label{fig:propertiesMP&Pom}
\end{figure}

In 4+1 dimensions, the ADM mass and the two angular momenta in
orthogonal planes of the Myers-Perry black hole \cite{MP} are     
\bea 
  M=\frac{3\,\pi \,\mu}{8\,G}\, ,  \hspace{8mm}
  J_{\phi}=\frac{\pi \, \mu\, a_1}{4\, G } \, , \hspace{8mm}
  J_{\psi}=\frac{\pi \, \mu\, a_2}{4\, G }\, . 
\eea
and the horizon area is
\be 
  A_\rom{H}=
  \sqrt{2}\, \pi^2 \,\mu 
  \left[ \mu - a_1^2 - a_2^2 
   + \big[ (\mu - a_1^2 - a_2^2)^2 - 4\, a_1^2 a_2^2 \big]^{1/2} \right]^{1/2}
\ee
Here $\mu$, $a_1$ and $a_2$  are the mass and
rotation parameters respectively. They must satisfy the condition
\bea \label{MPcond}
  \mu\geq a_1^2 +a_2^2 + 2 \,| a_1\, a_2| 
\eea
to ensure the
existence of an event horizon.

The single spinning Myers-Perry black hole shrinks to zero size while reaching
its maximum angular momentum at $j=1$, where the solution becomes
singular. This behaviour changes as the angular momentum in the orthogonal 
plane is turned on. An extremal non-zero minimum of the
horizon area is then reached at maximum momentum. This is shown in
figure \ref{fig:propertiesMP&Pom}. The endpoints of the fixed-$j_\phi$
curves all have zero temperature. The phase of 
extremal Myers-Perry black holes is shown as the gray curve in figure
\ref{fig:propertiesMP&Pom}. It is obtained by saturating the
bound \reef{MPcond}, which implies that the inner and outer
horizons coincide, so that the solution is extremal and has zero
temperature.


\end{document}